\documentclass[10pt,twocolumn,superscriptaddress]{revtex4-1}

\usepackage{amsmath, bm, amsfonts, amssymb}     
\usepackage{graphicx}                       
\usepackage{xfrac}                          
\usepackage{grffile}                        
\usepackage{color,colortbl}
\usepackage{xcolor}

\usepackage[colorlinks, citecolor=blue, urlcolor=red, linkcolor=blue]{hyperref}

\newcommand{\smf}[1] {\textcolor{blue}{\{\small{SMF: #1}\}}}

\newcommand{\be}{\begin{equation}}
\newcommand{\ee}{\end{equation}}
\newcommand{\bea}{\begin{eqnarray}}
\newcommand{\eea}{\end{eqnarray}}

\newcommand{\tw}{t_{\rm w}}
\newcommand{\ts}{t_{\rm s}}
\newcommand{\gdot}{\dot{\gamma}}
\newcommand{\sigmay}{\sigma_{\rm y}}

\newcommand{\kbt}{k_{\rm B}T}

\newcommand{\vecv}[1]{\bm{{#1}}}
\newcommand{\tens}[1]{\bm{{#1}}}

\newcommand{\xg}{x_{\rm g}}

\begin{document}

\title{Elastoviscoplastic rheology and ageing in a simplified soft glassy constitutive model}

\author{Suzanne M. Fielding}
 \affiliation{Department of Physics, Durham University, Science
 Laboratories, South Road, Durham DH1 3LE, UK}

\date{\today}

\begin{abstract}

Yield stress fluids display a rich rheological phenomenology. Beyond
the defining existence of a yield stress in the steady state flow
curve, this includes in many materials rather flat viscoelastic
spectra over many decades of frequency in small amplitude oscillatory
shear; slow stress relaxation following the sudden imposition of a small
shear strain; stress overshoot in shear startup; logarithmic or
sublinear power law creep following the imposition of a shear stress
below the yield stress; creep followed by yielding after the
imposition of a shear stress above the yield stress; richly featured
Lissajous-Bowditch curves in large amplitude oscillatory shear; a
Bauschinger effect, in which a material's effective yield strain is
lowered under straining in one direction, following a preceding strain
in the opposite direction; hysteresis in up-down shear rate sweeps;
and (in some materials) thixotropy and/or rheological ageing. A key
challenge is to develop a constitutive model that contains enough
underlying mesoscopic physics to have meaningful predictive power for
the full gamut of rheological behaviour just described, with only a
small number of model parameters, and yet is simple enough for use in
computational fluid dynamics to predict flows in complicated
geometries, or complicated flows that arise due to spontaneous
symmetry breaking instabilities even in simple geometries. Here we
introduce such a model, motivated by the widely used soft glassy
rheology model, and show that it captures all the above rheological
features.

\end{abstract}

\maketitle

\section{Introduction}
\label{sec:intro}


Many soft materials behave as so-called ``yield stress
fluids''~\cite{ISI:000407999000001,ISI:000419993800001,coussot2014yield,ISI:000266878700032,ISI:000246368400004,ISI:000293295000002,balmforth2014yielding,moller2006yield}.
Examples
include dense colloids, emulsions, foams, star polymers, microgels and
lamellar onion phases, as well as low density attractive colloidal
gels and clays. At imposed stresses below a critical yield stress,
$\sigma <
\sigmay$, they show solid-like rheological behaviour. In contrast, at larger  stresses, $\sigma>\sigmay$, they yield and flow like liquids.  Their steady state flow curve of shear stress $\sigma$ as a function of shear rate $\gdot$, typically measured in a slow shear rate sweep,
is then often fit to a Herschel-Bulkley form~\cite{ISI:000202025500001},
$\sigma(\gdot)=\sigmay+K\gdot^n$, with $0<n< 1$, or Bingham behaviour~\cite{bingham1922fluidity} with $n=1$.

In terms of the physical origin of this behaviour, yield stress fluids
can (broadly and loosely) be subdivided into two main categories. In
the first category, a material's constituent mesoscopic substructures,
such as colloidal particles, attract to form weakly flocculated
aggregates~\cite{larson2019review}. Even though the volume fraction of the
constituent particles can be quite low for the system overall, their
aggregates can result in a gel-like response at low loads, but are
then pulled apart and refluidised in shear.

In the second category, a material's constituent substructures are too
densely packed to properly rearrange at low loads, but do rearrange in
shear.  Examples include colloids, emulsions, foams, microgels, {\it
etc.}, which respectively comprise densely packed colloidal particles,
emulsion droplets, foam bubbles, or microgel beads. Materials in this
second category can be further divided into (at least) two idealised
limiting subcategories~\cite{ISI:000306323400007,ISI:000322230300001}:
thermal hard sphere colloids, in which the yield stress has a typical
magnitude $\kbt/R^3$, where $R$ is the particle radius, and athermal
soft suspensions, in which the yield stress has a typical magnitude
set by the modulus of the constituent particles. Materials in the second
subcategory have been termed ``soft glassy
materials''~\cite{ISI:A1997WM06400048,ISI:000085655200008,ISI:000074893400095}.
The constitutive model that we shall present in what follows is aimed
in particular at dense athermal soft particle suspensions, and
motivated by the original, widely used ``soft glassy rheology'' (SGR)
model~\cite{ISI:A1997WM06400048,ISI:000085655200008,ISI:000074893400095}.
It is worth noting, however, that the model actually also captures
many of the observed rheological features of dense hard sphere
colloids, and of low density attractive gels.

Beyond the defining presence of a yield stress in the steady state
flow curve, yield stress fluids also display a host of interesting
{\em dynamic} rheological behaviours. In linear response, the
viscoelastic spectra characterising their stress response to a small
amplitude oscillatory shear strain typically show a rather flat, power
law dependence over many decades of the oscillation frequency,
$\omega$~\cite{mackley1994rheological,ketz1988rheology,khan2000foam,mason1995elasticity,panizza1996viscoelasticity,hoffmann1993aggregating,mason1995linear}. The
storage modulus, $G'(\omega)$, typically exceeds the loss modulus,
$G''(\omega)$, by about an order of magnitude, consistent with a
nearly elastic response overall for these small deformations. The
presence of non-trivial dissipation (a non-zero $G''$) even at the
lowest frequencies accessible experimentally however also reveals a
broad underlying spectrum of sluggish stress relaxation modes. The
stress decay following the sudden imposition of a small shear strain
occurs over a similarly wide range of sluggish relaxation
timescales~\cite{ISI:000172642100029}.

The behaviour of yield stress fluids in time-dependent nonlinear flows
is similarly rich, in both strain-imposed and stress-imposed
protocols. In shear startup from rest, for example, they typically
display an initially elastic solid-like regime in which the stress
increases linearly with strain up to a maximum `overshoot'
value. Following this stress overshoot, the material yields and the
stress declines to its value in the ultimate fluidised flowing state,
prescribed by the steady state flow curve. Such behaviour has been
observed in foams~\cite{khan2000foam},
emulsions~\cite{batista2006colored,papenhuijzen1972role},
carbopol~\cite{divoux2010,ISI:000295085700080},
Laponite~\cite{Martinetal2012a,gibaud2008}, a fused silica
suspension~\cite{Kurokawa}, attractive
gels~\cite{koumakis2011two,liddel1996yield}, and waxy crude
oil~\cite{ISI:000341025500004}.

Following the imposition of a step shear stress below the yield
stress, a sustained solid-like slow creep response is typically seen,
in which the strain increases logarithmically or as a sublinear power
of time, with the strain rate accordingly decreasing as a power
law. In this way, the material creeps forward at an ever slowing rate,
but never attains a steady flow of non-zero rate. Following the
imposition of a shear stress just above the yield stress, in contrast,
an early time creep regime is followed at later times by a dynamical
yielding process in which the shear rate increases up to its final
value prescribed by the steady state flow curve. Such behaviour has
been observed in carbopol
gel~\cite{ISI:000294447600069,ISI:A1990DZ50400005}, carbon
black~\cite{ISI:000280140800011,ISI:000332461800012,sprakel2011stress},
polycrystalline hexagonal columnar phases~\cite{bauer2006collective},
and colloidal
glasses~\cite{ISI:000357577400001,siebenburger2012creep}.

Back-and-forth strain, strain rate or stress ramps or oscillations
have also been widely studied. In large amplitude oscillatory shear
(LAOS) experiments on yield stress
fluids~\cite{Yoshimura1987,Knaebel2000,Viasnoff2003,Rouyer2008,ISI:000273854300007,Renou2010,VanderVaart2013,Koumakis2013,Poulos2013,Poulos2015,ISI:000242219900019,ISI:000342206200016,ISI:000288162500028,ISI:000312240900002,ISI:000369750400007},
parametric Lissajous-Bowditch (LB) plots of stress versus strain
typically show a rather complicated progression in shape with
increasing amplitude of the applied shear. For example, characteristic
diamond shaped LB curves are often seen for intermediate
amplitudes. LB curves  have recently been modelled within the SGR model in Refs.~\cite{ISI:000438883200004,ISI:000427032600013,ISI:000390226100005}. Yield stress materials often also display a Bauschinger
effect~\cite{bannantine1990fundamentals,ISI:000280777900001}, in which
the apparent yield strain is reduced for straining in one direction,
following a preceding plastic strain in the opposite direction. The
stress response of yield stress fluids to shear rate sweeps often
displays a pronounced hysteresis between the downward and upward
sweeps, with the size of the hysteresis loop increasing with
increasing sweep rate~\cite{ISI:000396030800011,ISI:000313006100057}.

Indeed, the rheological response of many yield stress fluids also
shows a pronounced dependence on the `waiting time' since a sample was
prepared, before a flow is applied. This phenomenon is often termed
rheological ageing and/or thixotropy. For a precise definition of
rheological ageing, see Ref.~\cite{ISI:000085655200008}. The
definition of thixotropy and its distinction from viscoelasticity and
from ageing is a topic of ongoing discussion~\cite{larson2019review}.
Typically, a sample that has waited longer in an undisturbed state
before a flow is applied will show a higher viscosity and/or a more
solid-like response. The latter can be evidenced, for example, by a
slower relaxation of stress following the rapid imposition of a small
shear strain, or a larger stress overshoot in shear startup in older
samples. Typically, a material is then rejuvenated to a state of lower
viscosity and/or lower solidity by an imposed flow.

During the dynamical process whereby a material yields from an
initially solid-like to finally fluid-like state, a state of initially
homogeneous shear will often become unstable to the formation of
heterogeneous shear bands. This has been observed experimentally in
yield stress fluids during shear
startup~\cite{divoux2010,ISI:000295085700080,Martinetal2012a,gibaud2008,Kurokawa,ISI:000341025500004},
step
stress~\cite{ISI:000294447600069,ISI:A1990DZ50400005,ISI:000280140800011,ISI:000332461800012,ISI:000357577400001},
flow curve sweeps~\cite{ISI:000313006100057}, and
LAOS~\cite{ISI:000242219900019,ISI:000342206200016,ISI:000369750400007,ISI:000288162500028}.
It has also been studied theoretically and computationally in these
same protocols of shear
startup~\cite{ISI:000342206200002,ISI:000386386400004,ISI:000246210200042,varnik-jcp-120-2788-2004,ISI:000250675300003,ISI:000278158800014,ISI:000286879900011,ISI:000344142000001,ISI:000322544200022,PhysRevE.76.056106,Manningetal2009a,ISI:000285583500029},
step
stress~\cite{ISI:000325376600001,ISI:000315141600016,ISI:000344142000001},
flow curve sweeps~\cite{ISI:000396030800011}, and
LAOS~\cite{ISI:000427032600013,ISI:000390226100005}. In many
materials, the shear bands that form during yielding then gradually
heal away to leave a homogeneous steady flowing state. Some yield
stress fluids instead support shear bands in the ultimate steady
flowing state~\cite{Coussotetal2002c,ISI:000242151800010}. We do not
consider such materials here: the model that we discuss has a
monotonically increasing constitutive relation between stress and
strain-rate, precluding steady state shear banding.

From a theoretical viewpoint, a key challenge is to understand how the
macroscopic flow properties just described emerge out of the
underlying collective dynamics of a material's constituent mesoscopic
substructures, for any given category of yield stress fluids, and to
build this understanding into a rheological constitutive
model. Ideally, such a model should contain enough of the key
mesoscopic physics to have meaningful predictive power for the full gamut
of observed rheological phenomena, with just a small number of model
parameters. At the same time, it should also be simple enough for use
in computational fluid dynamics (CFD) to address flows in complicated
geometries, or complicated flows that arise via spontaneous symmetry
breaking instabilities even in simple geometries. It should therefore
preferably be of time-differential form, which is much easier to
implement numerically in a CFD solver than a model of time-integral
form. The primary contribution of this work is to introduce a
constitutive model that for the first time, to this author's
knowledge, meets all these desirable criteria.

We start in Sec.~\ref{sec:review} by briefly reviewing some of the
most widely used models of yield stress rheology in the existing
literature. In Sec.~\ref{sec:SGR} we discuss one such model in more
detail: the soft glassy rheology (SGR) model. This does capture all
the rheological phenomena described above, but is in its present form
far too complicated for use in CFD. Indeed, even computations of
homogeneous simple shear flows can be very cumbersome within the SGR
model its existing form. Accordingly, in Sec.~\ref{sec:SGRsimple} we
introduce a simplified SGR model, and in Sec.~\ref{sec:results}
demonstrate it to indeed capture all the rheological features
discussed above. The potential contributions of this new model are
twofold. First, it will allow significantly more straightforward
computation of homogeneous shear flows for anyone wishing to fit the
SGR model's predictions to rheometric data. Second, it renders SGR
feasible for use in CFD, once suitably tensorialised. We therefore
suggest a possible generalisation to tensorial stresses in
Sec.~\ref{sec:tensorial}, before finally setting out our conclusions
in Sec.~\ref{sec:conclusions}.

\section{Overview of existing constitutive models}
\label{sec:review}

Existing elastoviscoplastic constitutive models range from those built
from bottom up on the basis of microscopic or mesoscopic physics, to
those posed from top down on the basis of macroscopic
phenomenology. We now summarise some of the most widely used models in
the existing literature, and the extent to which they meet the
desirable criteria set out in the penultimate paragraph of
Sec.~\ref{sec:intro} above.

\subsection{Microscopically derived models}

For dense colloidal suspensions, a rheological constitutive theory has
been built on the mode coupling theory (MCT) of the colloidal glass
transition~\cite{ISI:000259680600077,ISI:000312240900007}. It starts
by writing an equation of motion for the microscopic probability
distribution in configuration space of the position vectors of a dense
ensemble of Brownian particles (ignoring hydrodynamic
interactions). This microscopic equation is then projected via a
series of approximations onto a time-integral rheological constitutive
equation for macroscopic stresses. This takes as its basic input the
material's static and dynamic structure factors for the underlying
microscopic density correlation functions.

MCT successfully captures many of the observed rheological features of
dense colloidal suspensions, including the existence of a yield stress
in the flow curve. Its formalism is heavy to implement in
computational practice, however, even in simple homogeneous shear
flows. Nonetheless, very recent work has incorporated a simplified
schematic -- although still time-integral -- MCT constitutive equation
into a lattice Boltzmann solver for CFD in channel shear flow,
assuming translational invariance in the flow
direction~\cite{ISI:000366319700022}.

\subsection{Mesoscopic elastoplastic models}

Mesoscopic elastoplastic models conceptually divide a macroscopic
sample of material into many local mesoscopic regions, each of which
is ascribed continuum variables of local strain and stress relative to
a locally undisturbed equilibrium. Each such region is represented as
an elastoplastic element that loads elastically in flow up to a local
threshold, after which it yields plastically, then assumes a new
elastic state relative to a new locally undeformed equilibrium.


In lattice-based elastoplastic models~\cite{nicolas2018deformation},
the elements just described reside on a lattice, and the stress
relaxation involved in any local plastic yielding event results in an
explicit redistribution of stress to surrounding elements via an
Eshelby propagator, ensuring that force balance is properly
maintained~\cite{ISI:000226675600003}. Such models capture many
observed features of elastoplastic rheology (even though they often
fail properly to account for the advection of an element's position in
flow). In containing detailed spatial information about stress
propagation, however, they are too computationally intensive in their
present form for use in CFD to predict flows in anything other than
small and simply shaped geometries.


Mean field elastoplastic models instead discard any explicit spatial
information about the location of elements, and the stress propagation
that follows local yielding events. Instead, they model stress
propagation in a mean field way.  For example, the Hebraud-Lequeux
model does so by invoking a diffusive term in the equation of motion
for the probability distribution of local strains, with a diffusion
constant set by the sample-average yielding
rate~\cite{hebraud1998mode}. Mean field models with broader-tailed
stress propagation statistics have also been
studied~\cite{ISI:000368519600002}. Most such models, and their
lattice-based counterparts described above, assume a flat distribution
of local yield energy thresholds.

The soft glassy rheology (SGR)
model~\cite{ISI:A1997WM06400048,ISI:000085655200008,ISI:000074893400095}
instead assumes a distribution of local yield energy thresholds with
an exponential tail. It furthermore models stress propagation by means
of an effective temperature that can activate any element out of its
local energy minimum and thereby trigger a local yielding event. This
activation is taken to model, in a mean field way, stress propagation
from other local yielding events elsewhere in the sample. Starting
from the initial purely mean field model, SGR was later extended to
address flows that develop spatial structuring in one spatial
dimension, either via shear banding~\cite{ISI:000266798200007} or
extensional necking~\cite{ISI:000352990500012}, by allowing stress
propagation in the relevant dimension.

Although the mean field elastoplastic models just described are simpler
than their lattice-based counterparts, they still in general involve
the time evolution of a full distribution of local strain variables,
and remain as yet too complicated to implement in CFD.

An alternative mesoscopic approach, originally intended to model the
deformation properties of metallic glasses, is based on the collective
statistics of many `shear transformation zones' (STZs), which resemble
the yielding local elements of the elastoplastic models just
described~\cite{PhysRevE.76.056106,Manningetal2009a}.

\subsection{Phenomenological macroscopic models}

Besides the microscopic and mesoscopic models just described, other
constitutive models of yield stress rheology have been built from the
top down, on the basis of macroscopic phenomenology.  The earliest
such models posited a static relation between stress and strain rate
~\cite{ISI:A1987H891000002,ISI:000337874400008,ISI:000173617600006,ISI:A1985ATS6200011,ISI:000171099500010,ISI:A1997XB14900003}.
When used in CFD, however, these necessitate a cumbersome separate
calculation of the `yield surface', or regularisation of sub-yield
behaviour~\cite{ISI:000398072300007}.  They also miss most of the key
physics, including all the dynamical rheological phenomena summarised
in Sec.~\ref{sec:intro} above.

More recent phenomenological models therefore instead posit an
evolution equation for the stress, in order to account for the stress
in a material at any time as a functional of the strain rate {\em
history} it has experienced.  This equation may depend on one or more
auxiliary variables, for which evolution equations are also posited.
Examples include fractional
calculus~\cite{ISI:000406087600005,ISI:000345194700006}, structural
evolution, fluidity, and elastoviscoplastic
models~\cite{ISI:000312240900002,ISI:000341025500004,ISI:000419395900024,ISI:000388430800020,ISI:000410747200004,ISI:000249049200001,ISI:000288162500032,ISI:000265317500017}. Although
vastly superior to the static models, many involve 10-20+ fitting
parameters, limiting their predictive power. Indeed, models
benchmarked by fitting to straightforward strain-imposed protocols
such as shear startup often then perform poorly in more complicated
oscillatory/reversal protocols, and/or in stress-imposed
protocols. Such models furthermore often incorporate phenomenological
notions such as those of a `back stress' or `kinematic hardening',
without always offering a clear understanding of such concepts in
terms of the underlying microscopic or mesoscopic physics.

\subsection{Summary of existing models}

Among the constitutive models just described, those based on
underlying microscopic and mesoscopic physics tend to perform best at
predicting the broad gamut of dynamical rheological phenomena
described above, but are often prohibitively complicated for use in
CFD to predict macroscopic flows in complex geometries.  In contrast,
the macroscopic phenomenological models are generally better suited to
CFD, but often contain many model parameters, and/or capture only a
subset of the desired rheological phenomenology, and/or are limited in
the underlying physics they contain.

Indeed, to this author's knowledge, no currently existing constitutive
model of elastoviscoplastic yield stress rheology currently exists
that satisfies all the desirable criteria set out above: of containing
enough underlying micro/mesoscopic physics to predict the rich
dynamical rheology of yield stress fluids with just a small number of
model parameters, while also being simple enough -- and preferably of
time-differential form -- for use in CFD to predict flows in
complicated geometries.

\section{Original soft glassy rheology model}
\label{sec:SGR}

The soft glassy rheology (SGR)
model~\cite{ISI:A1997WM06400048,ISI:000085655200008,ISI:000074893400095}
considers an ensemble of elastoplastic elements, each representing a
local mesoscopic region of soft glassy material (a few tens of
emulsion droplets, say). Each element is assigned local continuum
variables of shear strain $l$ and shear stress $kl$, describing the
mesoscopic region's state of elastic deformation relative to a locally
undeformed equilibrium.  In between local yielding events, the strain
of each element affinely follows the macroscopic shear,
$\dot{l}=\gdot$, giving an elastic buildup of stress.

The stress is intermittently released by local plastic yielding
events. In any such event, a mesoscopic region suddenly rearranges
into a new configuration locally. This is modelled by its
representative element hopping between traps in an energy
landscape. An element in a trap of depth $E$ and with local shear
strain $l$ is assigned a probability per unit time of hopping of
$\tau^{-1}(E,l)$, with
\be
\tau(E,l)=\tau_0\exp\left[(E-\tfrac{1}{2}kl^2)/x\right].
\ee
The stored elastic energy $\tfrac{1}{2}kl^2$ at any instant therefore
offsets the bare trap depth $E$, leading to a reduced local barrier to
rearrangement, $E-\tfrac{1}{2}kl^2$. This confers rheological shear
thinning on the sample as a whole. After hopping, an element selects a
new trap depth at random from a prior distribution
\be
\label{eqn:postHop}
\rho(E)=\frac{1}{\xg}\exp\left(-E/\xg\right),
\ee
and resets its local strain $l$ to zero, thereby relaxing the local
stress.

With the dynamics just described, the probability $P(E,l,t)$ for an
element to be in a trap of depth $E$ with local shear strain $l$
evolves according to
\be
\label{eqn:master}
\dot{P}(E,l,t)+\gdot\frac{\partial P}{\partial l} = -\frac{1}{\tau(E,l)}P+Y(t)\rho(E)\delta(l).
\ee
The advected derivative on the left hand side captures the affine
loading of each element by shear. The first (`death') term on the
right hand side describes hops out of traps. The second (`birth') term
describes hops into the bottom of traps, $l=0$, with the new trap
depth chosen at random from the prior distribution, $\rho(E)$, and with
an ensemble average hopping rate
\be
Y(t)=\int dE \int dl \frac{1}{\tau(E,l)}P(E,l,t).
\ee

The macroscopic stress of the sample as a whole is defined as the
average over the local elemental ones:
\be
\sigma(t)=k\int dE \int dl\; l P(E,l,t).
\ee

Combined with the exponential prior, $\rho(E)$, the exponential
activation factor $\tau(E,l)$ confers a glass transition at a noise
temperature $x=\xg$.  In the absence of any applied flow, the model
displays rheological ageing in the glass
phase~\cite{ISI:000085655200008}, $x<\xg$: following sample preparation
at time $t=0$ by means of a sudden quench from a high initial noise
temperature to a value $x<\xg$, the system slowly evolves into ever
deeper traps as a function of time $t$. This results in a growing
stress relaxation time, $\langle
\tau\rangle\sim t$, and therefore in ever more solid-like rheological response as the sample ages.  An imposed shear of constant rate
$\gdot$ will however arrest ageing and rejuvenate the sample to a
steady flowing state of effective age $\langle \tau\rangle\sim
1/\gdot$. The steady state flow curve has a yield stress $\sigmay$
that grows linearly with $\xg-x$ in the glass phase $x<\xg$.

For many soft glassy materials, the typical energy barrier for
rearrangements greatly exceeds thermal energies. Accordingly, the
parameter $x$ is not the true thermodynamic temperature but is taken as
an effective noise temperature that models in a mean field way
coupling with other yielding events elsewhere in the sample.

As noted above, the SGR model captures a glass transition and ageing
by invoking the exponential form for the probability distribution of
post-hop yield-energies in Eqn.~\ref{eqn:postHop}. Recent particle
based simulations explicitly measured the distribution of yield
thresholds and local stresses for a sheared Lennard-Jones
glass~\cite{ISI:000380122800005,ISI:000426631500004,RottlerPreprint}. See
also Ref.~\cite{SollichCecam}. It would clearly be interesting to
undertake such a study for an athermal suspension of purely repulsive
soft particles.

The full soft glassy rheology model just described captures many
features of the elastoplastic rheology of yield stress fluids. These
include a yield stress in the steady state flow curve; broad and
rather flat power-law viscoelastic spectra; ageing in the power-law
stress decay following a small amplitude step shear strain; an
overshoot in the shear stress following the switch-on of a shear of
constant rate $\gdot$; slow creep during which the shear rate
decreases as a power law $\gdot\sim t^{-\alpha}$ after the imposition
of a shear stress below the yield stress; slow creep over several time
decades followed by yielding and a sudden increase of the shear rate
to a steady flowing state after the imposition of a shear stress just
above the yield stress; and characteristic diamond or rhomboidal
shaped Lissajous curves in large amplitude oscillatory shear strain.


The ability of the SGR model to capture this rich rheological
phenomenology within such a minimal and powerfully generic set of
physical assumptions, and with just one free parameter - the noise
temperature, $x$ - is a remarkable achievement. Set against that
achievement, this simplicity at the same time also inevitably limits
the model's ability to fit experimental data at a detailed {\em
quantitative} level. As such, a worthwhile avenue for future study
could be to bring in additional microscopic physics characterised by a
small number of additional model parameters, potentially opening up
the possibility of quantitatively fitting experimental data, while
still keeping the number of model parameters and assumptions much
smaller than in existing top-down phenomenological models. One such
approach could be to build deformation and flow into a multi-layer
glassy trap model along the lines of that in
Ref.~\cite{bouchaud1995aging}.

We have described here the SGR model in its original form, which
considers only scalarised shear stresses, and which contains no
spatial information about the location of any element. As such, it
addresses only homogeneous simple shear flows. Extensions of the model
have since been put forward to address tensorial
stresses~\cite{ISI:000187952400012}, and flows that become
heterogeneous in one spatial dimension due to shear
banding~\cite{ISI:000266798200007} or extensional
necking~\cite{ISI:000352990500012}.
 
\section{Simplified soft glassy rheology model}
\label{sec:SGRsimple}

\subsection{Motivation for a simplified model}

As noted above, the fact that the SGR model captures the rich
phenomenology just described within a relatively simple and generic
set of physical assumptions, and with a small number of model
parameters, is a remarkable achievement. Indeed, once suitable units
have been chosen, the model's only free parameter is the noise
temperature, $x$.

Set against this appeal is the considerably cumbersome task of
computing these rheological behaviours in practice. Even for
homogeneous simple shear flows, this requires the solution either of
the full partial differential equation (PDE) given above, $\partial_t
P(E,l,t)=\cdots$, or the solution of two coupled nonlinear integral
constitutive equations with power-law memory kernels (derived from the
original PDE)~\cite{ISI:000074893400095}, or the direct simulation of
hopping SGR elements~\cite{ISI:000390226100005}, typically with $10^5$
elements required for reliable predictions. (This direct simulation is
generally easier to implement computationally than a numerical
solution of the differential or integral equations, but is still
costly in terms of computer time.)

To utilise the (tensorially extended) full SGR model in computational
fluid dynamics (CFD) to address heterogeneous flows in complicated
geometries would require comparably involved computation at each
lattice site: a task that is likely to prove prohibitively formidable
in both practical implementation and computational cost. Indeed, to
this author's knowledge, such a task has not been attempted to date.

The present manuscript therefore aims to develop a simplified SGR
model that captures the same rich phenomenology as the original model,
but with greatly reduced computational demand. The contributions of
this will be twofold. First, the calculation of homogeneous flows will
be made simpler for anyone wishing to compare rheometric experimental
data with SGR.  Second, the SGR model will be rendered simple enough
for practical use in CFD for predicting elastoplastic flows in
complicated geometries, or complicated flow patterns that arise via
spontaneous symmetry breaking instabilities even in simple geometries.

We shall undertake this simplification first in the context of a
scalarised approach that considers only shear strains and stresses,
before returning in Sec.~\ref{sec:tensorial} below to suggest a tensorial
generalisation, as needed for CFD.

\subsection{Simplified model}

We start by exactly rewriting the full SGR model equation, Eqn.~\ref{eqn:master}, as follows: 
\be
\label{eqn:master1}
\dot{P}(E,l,t)+\gdot\frac{\partial P}{\partial l} = -\frac{1}{\tilde{\tau}(E)}f(l)P+Y(t)\rho(E)\delta(l),
\ee
in which we have written the `bare' hopping time
\be
\tilde{\tau}(E)=\tau_0\exp\left(\frac{E}{x}\right),
\ee
and the `boost factor' to the hopping rate:
\be
f(l)=\exp\left(\frac{kl^2}{2x}\right).
\ee
We now exactly rewrite the full joint probability distribution $P(E,l,t)$ of
an element being in a trap of depth $E$ with a local strain $l$ as the
probability $G(E,t)$ of an element being in a trap of depth $E$
multiplied by the conditional probability $P_1(l|E,t)$ of an element
having a local strain $l$, given that it is in a trap of depth $E$:
\be
P(E,l,t)=G(E,t)P_1(l|E,t).
\ee
We have suggestively used the notation $G(E,t)$ rather than
$P(E,t)$ in the first term on the right hand side, because this
quantity will below assume the behaviour of a modulus-like quantity
(once re-dimensionalised by $k$).

We can then exactly rewrite Eqn.~\ref{eqn:master1} as
\be
\label{eqn:fullHere}
\dot{G}(E,t)P_1 + G\dot{P_1}+\gdot G\frac{\partial P_1}{\partial l}=-\frac{G}{\tilde{\tau}(E)}f(l)P_1 + Y(t)\rho(E)  \delta(l).
\ee
Averaging this equation over $l$ at fixed $E$ then gives
\be
\label{eqn:oneExact}
\frac{d}{dt}G(E,t)=-\frac{G(E,t)}{\tilde{\tau}(E)}\overline{f(l)}(E,t)+Y(t)\rho(E).
\ee
Instead pre-multiplying Eqn.~\ref{eqn:fullHere} by $l$ before averaging  over $l$ at fixed $E$ gives the additional equation:
\be
\label{eqn:twoExact}
\frac{d}{dt}\left[G(E,t)\bar{l}(E,t)\right]=G(E,t)\gdot-\frac{G(E,t)}{\tilde\tau(E)}\overline{l f(l)}(E,t).
\ee
In these equations, we have used the notation
\be
\overline{a(l)}(E,t)=\int dl\, P_1(l|E,t) a(l).
\ee
for any function $a(l)$.

Eqns.~\ref{eqn:oneExact} and~\ref{eqn:twoExact} together constitute an
{\em exact} rewriting of the full SGR model. We now make an {\em
approximation} by rewriting
\be
\overline{a(l)}=a(\bar{l}).
\ee
This amounts to assuming that the distribution $P_1(l|E,t)$ has the
form of a delta function located at $\overline{l}(E,t)$: {\it i.e.},
that traps of depth $E$ have a single slaved local strain,
$\overline{l}(E,t)$. For simplicity we further now drop the overbar
notation from $l$. Eqns.~\ref{eqn:oneExact} and~\ref{eqn:twoExact} can
then be written:
\be
\label{eqn:oneApprox}
\dot G(E,t)=-\frac{G(E,t)}{\tau(E,l(E,t))}+Y(t)\rho(E),
\ee
and
\be
\label{eqn:twoApprox}
\dot\sigma(E,t)=kG(E,t)\gdot -\frac{\sigma(E,t)}{\tau(E,l(E,t))}.
\ee
Here we premultiplied Eqn.~\ref{eqn:twoExact} by $k$ and have defined the stress in traps of depth $E$,
\be
\sigma(E,t)=kG(E,t)l(E,t).
\ee 
The average hopping rate,
\be 
\label{eqn:integralOne}
Y(t)=\int dE\frac{G(E,t)}{\tau(E,l(E,t))},
\ee 
with an $E-$dependent relaxation timescale 
\be
\tau(E,l(E,t))=\tau_0 \exp\left[(E-\tfrac{1}{2}kl(E,t)^2)/x\right].  
\ee 
Normalisation of overall element numbers demands that
\be
\label{eqn:integralTwo}
\int dE\,G(E,t)=1.
\ee
As above, the prior distribution is
\be
\rho(E)=\frac{1}{\xg}\exp(-E/\xg).
\ee 

Throughout what follows, in both the original and simplified SGR
models, we shall choose units of time in which $\tau_0=1$ and of energy in
which $\xg=1$. We further rescale strains such that $k=1$, making the
typical yield strain of order unity.

\subsection{Discussion of the simplified model}

The simplified SGR model just described has a rather appealing
physical structure. Indeed, for any fixed value of energy depth $E$,
Eqn.~\ref{eqn:twoApprox} takes the form of a Maxwell model in which an
elastic loading term with an effective modulus $G(E,l)$ (in our units)
competes with a plastic stress relaxation term. The relaxation time
$\tau(E,l(E))$ is however a strongly nonlinear function of the energy
depth $E$ and local strain $l$. The modulus, $G(E,t)$, which is
prescribed by the fraction of elements in traps of depth $E$,
furthermore has its own dynamics given by
Eqn.~\ref{eqn:oneApprox}.

In assuming a single value of the local strain $l(E,t)$ for all
elements in traps of a given energy depth $E$, we have reduced the
partial differential equation of the full SGR model, $\partial_t
P(E,l,t)=\cdots$, with two dynamical variables, to two equations,
$\partial_t G(E,t)=\cdots$ and $\partial_t
\sigma(E,t)=\cdots$, each with just one dynamical variable. Neither  equation now contains any derivatives with respect to $E$, further simplifying any numerics.

Nonetheless, one must still in principle evolve the two full functions
$G(E,t)$ and $l(E,t)$ over time $t$. A second, pragmatic
simplification however arises in recognising that the continuous
spectrum of energy values can be discretised on a grid of $N$
values linearly distributed in a suitably chosen range $0
\le E \le E_{\rm max}$. This leaves  $2N$ 
differential equations, which, apart from the integral couplings of
Eqn.~\ref{eqn:integralOne} and~\ref{eqn:integralTwo}, are `ordinary'
in form.

Numerical results obtained within this simplified model then in
principle need converging to the limit $E_{\rm max}\to\infty,
N\to\infty$. Once converged, the results show excellent agreement with
all the qualitative predictions of the full SGR model, in every
rheological protocol studied. Indeed, they show exact quantitative
agreement in the regime of linear rheology. As we shall see below,
however, the quantitative numbers can differ from full SGR typically
by a worst-case factor of about 2 in the most nonlinear regime of
protocols such as those in Figs.~\ref{fig:flowCurves}
and~\ref{fig:startup}.

Given this quantitative discrepancy from full SGR, together with the
fact that even full SGR, with its minimal -- and therefore powerfully
generic -- set of physical assumptions, is anyway not expected to
model any particular experimental sample in a fully quantitative way,
we follow in our numerical computations below the pragmatic philosophy
of taking the minimal value of $N$ required to give convergence to the
$N\to
\infty$ limit of the simplified  model to within $1\%$. Typically,
we find $E_{\rm max}=12.0$ and $N=32$ sufficient for most
protocols. Indeed, even $N=16$ gives convergence to $5\%$, but we
take $N=32$ minimally in what follows.

This brings the number of degrees of freedom required to predict
homogeneous simple shear flows into the numerically trivial and
represents a considerable simplification compared with the full SGR
model, which required the solution of the full partial differential
equation $\partial_t P(E,l,t)=\cdots$, or the solution of two coupled
nonlinear integral constitutive equations with power-law memory
kernels, or the direct simulation of typically $10^5$ hopping SGR
elements. In performing CFD with a fully tensorial stress, a
corresponding number $7N$ of degrees of freedom would be needed at
each lattice site, as discussed further in Sec.~\ref{sec:tensorial}
below. This should prove feasible for computing 2D flow fields on a
small cluster. Even 3D flows should be feasible with further
parallelisation.
 
Finally, we reiterate that the assumption of a single value of $l$ for
any fixed $E$ represents an approximation that, as discussed above,
leads to typical factors of 2 between the quantitative predictions of
the full and simplified SGR models, in the most nonlinear regimes of
many protocols. Importantly, however, this assumption is in fact {\em
exactly correct in linear rheology}. The simplified SGR model derived
above is therefore predicted to give results that {\em exactly and
quantitatively} correspond to full SGR in any linear rheological
protocol. Our numerical results will indeed confirm this.

\begin{figure}[!t]
\includegraphics[width=1.0\columnwidth]{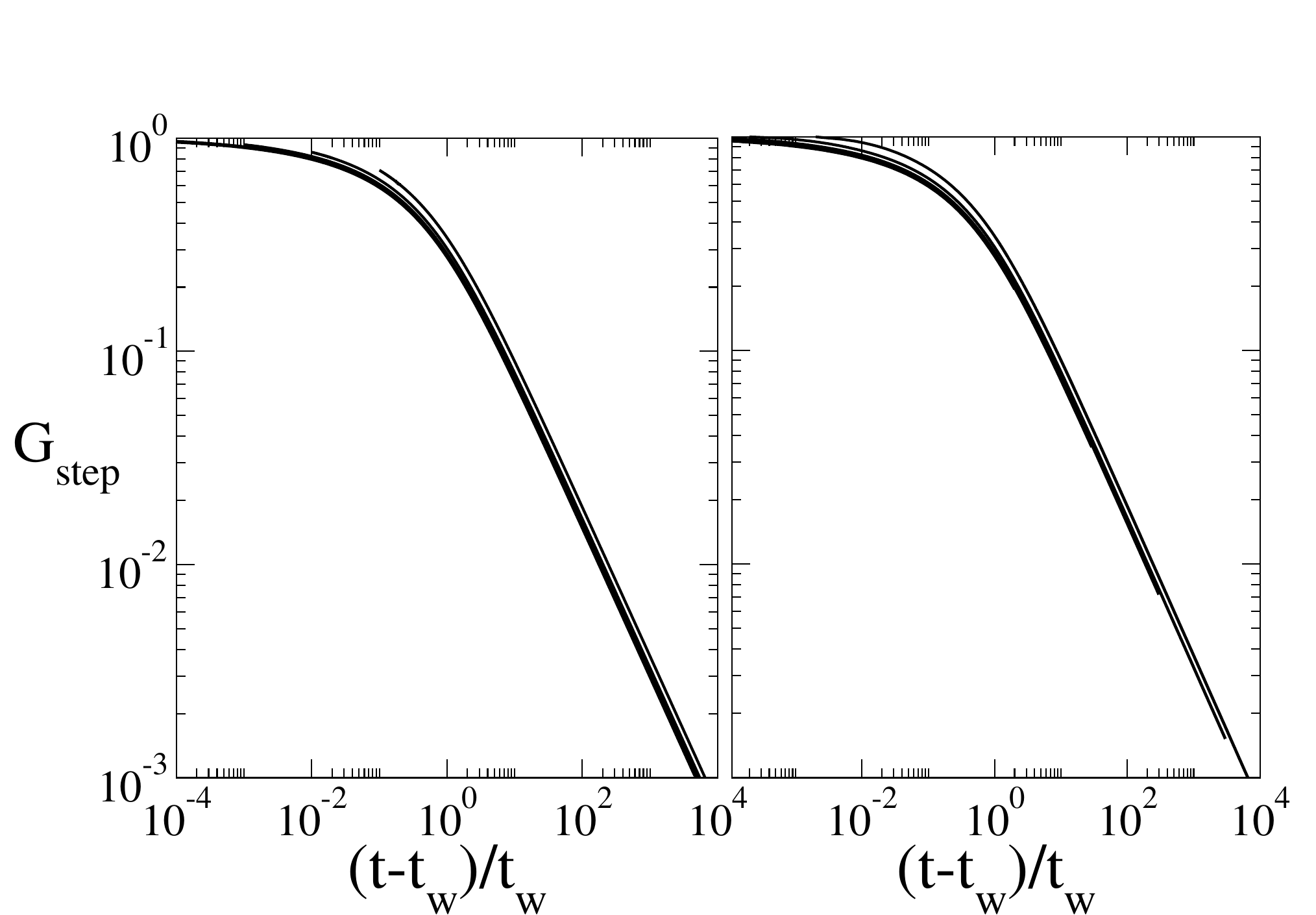}
\caption{Stress decay following a small amplitude step strain imposed at a waiting time $\tw$. {\bf Left:} results obtained within the full SGR model, by solving the integral constitutive equation of Ref.~\cite{ISI:000074893400095}. {\bf Right:} simplified SGR model. In both cases, the effective temperature $x = 0.7$, imposed shear strain $\gamma_0 = 0.001$, initial sample ages $\tw = 10^2, 10^3, 10^4, 10^5, 10^6$. In the simplified SGR model, $E_{\rm max} = 24.0$, $N = 32$, numerical timestep $dt=0.1$.}
\label{fig:stepStrain}
\end{figure}

With the exception of the right panel of Fig.~\ref{fig:viscospectra},
all the results presented will be in the model's glass phase,
$x<1$. Because the simplified model presented so far assumes a
scalarised shear stress, and contains no spatial information, all our
numerical predictions will pertain to a homogeneous simple shear
flow. In many of the protocols considered, however, shear bands would
be expected to arise in any model that allowed heterogeneous flows, as
described in Sec.~\ref{sec:intro} above. This would modify the
rheological signals to some extent compared with those computed below
within the assumption of a homogeneous flow.
 
\section{Rheological predictions of the simplified  SGR model}
\label{sec:results}

We now present our numerical results for the predictions of the
simplified SGR model in homogeneous simple shear flow. We start with
linear rheology in Sec.~\ref{sec:linear}, before turning to address
nonlinear flows. For any protocol in which the sample age explicitly
features, we model sample preparation at time $t=0$ via a sudden
quench from an infinite initial noise temperature to a final noise
temperature, usually (as just noted) in the glass phase, $x<1$. This
gives an initial distribution of trap depths $G(E,t=0)=\rho(E)$. We
further assume all local strains and stresses to be zero in this
initial state, corresponding to an initially well relaxed sample. We
then age the sample undisturbed for a time $\tw$, before imposing a
strain or stress according to the protocol in question.

The aim of this work is not to undertake direct fitting of the model's
predictions against any particular set of experimental data; nor yet
exhaustively to review the available data for each protocol that
follows. Nonetheless, we pause before presenting our results to
collect a few experimental references in which data qualitatively
according with the key features of several of the figures that follow
can be found: Fig.~\ref{fig:stepStrain}~\cite{ISI:000172642100029},
Fig.~\ref{fig:viscospectra}~\cite{mason1995elasticity,purnomo2008glass},
Fig.~\ref{fig:flowCurves}~\cite{mason1995elasticity},
Fig.~\ref{fig:startup}~\cite{ISI:000341025500004},
Fig.~\ref{fig:sweep}~\cite{ISI:000396030800011},
Fig.~\ref{fig:lissajous}~\cite{ISI:000341025500004},
Fig.~\ref{fig:creep}~\cite{siebenburger2012creep}. Indeed, Ref.~\cite{purnomo2008glass} carried out a quantitative fitting of the SGR model to measured viscoelastic spectra.

\subsection{Linear rheology}
\label{sec:linear}

\subsubsection{Stress relaxation after linear step shear strain}

A standard rheological test consists of suddenly straining a
previously undeformed material by an amount $\gamma_0$ at a time
$\tw$. The shear strain is accordingly
$\gamma(t)=\gamma_0\Theta(t-\tw)$, where $\Theta$ is the Heaviside
function. The shear stress response can be written generally as
\be
\label{eqn:Gstep}
\sigma(t)=\gamma_0 G_{\rm step}(t-\tw,\tw;\gamma_0).
\ee
In the limit of linear response, $\gamma_0\to 0$, the $\gamma_0$
dependence disappears from the stress relaxation function, $G_{\rm step}$. 

The left panel of Fig.~\ref{fig:stepStrain} shows results for
$G_{\rm step}(t-\tw,\tw)$ in this linear regime, computed within the full SGR
model, for several samples ages $\tw$ at a fixed noise temperature in
the glass phase. As can be seen, the model predicts slow power law
stress relaxation. It furthermore captures rheological ageing, in which
this stress relaxation takes place on a typical timescale that grows
as the age of the sample $\tw$. This gives the observed collapse of
the data for different values of $\tw$, as a function of the rescaled
time interval, $(t-\tw)/\tw$.

Corresponding results for the simplified SGR model are shown in the
right panel of Fig.~\ref{fig:stepStrain}, for matched parameter
values. Excellent agreement is obtained between the full and
simplified models, consistent with our above statement that the
assumption made in moving from the full to the simplified model is
exactly correct in the linear rheological regime.

\subsubsection{Viscoelastic spectra}

We now consider the viscoelastic spectra that characterise a
material's stress response to a small amplitude oscillatory shear
strain. As discussed in Ref.~\cite{ISI:000085655200008}, the
definition of viscoelastic spectra in an ageing material needs some
care, because the time-translational invariance (TTI) that is usually
implicitly assumed in defining these spectra breaks down as a
consequence of ageing.

\begin{figure}[!t]
\includegraphics[width=1.0\columnwidth]{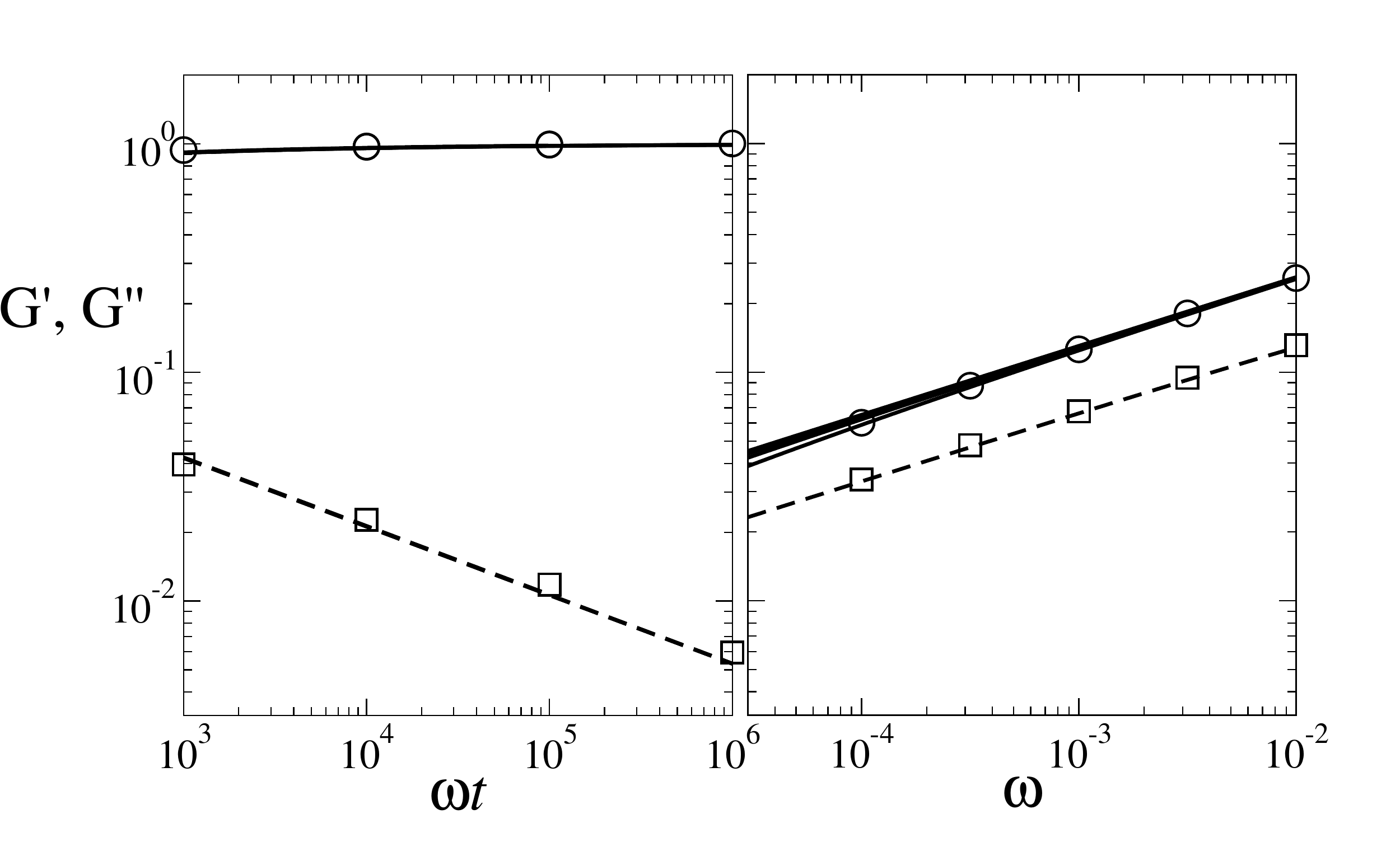}
\caption{Viscoelastic spectra of the full SGR model, obtained by solving the integral constitutive equation of Ref.~\cite{ISI:000074893400095} (solid lines: $G'$, dashed lines: $G''$) and simplified SGR model (circles: $G'$, squares: $G''$). {\bf Left:} for an effective temperature $x=0.7$, as a function of frequency scaled by system's age, $\omega t$. {\bf Right:} for an effective temperature $x=1.3$, as a function of bare frequency $\omega$.  Sample age $t=10^7$. In the simplified SGR model, $E_{\rm max}=24.0$, $N=64$, $dt=0.01$.}
\label{fig:viscospectra}
\end{figure}

Consider an experiment in which a sample is freshly prepared at
time $t=0$ then allowed to age undisturbed to a time $\ts$. A small
amplitude oscillatory shear strain of amplitude $\gamma_0$ is started
at this time $\ts$, and maintained up to a later time $t$. For such a
protocol, one can unambiguously define a {\rm time-dependent}
viscoelastic spectrum:
\bea
G^*(\omega,t,\ts)&=&i\omega\int_{\ts}^t\,dt'e^{-i\omega(t-t')}G_{\rm step}(t-t',t')\nonumber\\
                 & & + e^{-i\omega(t-\ts)}G_{\rm step}(t-\ts,\ts),
\eea
where $G_{\rm step}(t-t',t')$ is the (non-TTI) stress relaxation
function defined in Eqn.~\ref{eqn:Gstep} above, in the limit
$\gamma_0\to 0$.  In Ref.~\cite{ISI:000085655200008}, it was shown
that the dependence of $G^*$ on $\ts$ becomes negligible in full SGR
once many cycles have been performed, $\omega(t-\ts)\gg 1$, giving
$G^*(\omega,t,\ts)\to G^*(\omega,t)$. Although intuitively reasonable,
this simplification is not in fact guaranteed upfront in a glassy
material with long term memory. Nonetheless, we now adopt
$G^*(\omega,t)$ as a working definition of the time-dependent
viscoelastic spectrum for an ageing material. Note therefore that the current time $t$ plays the role of the sample age in this protocol, not the time $\ts$, now forgotten, at which the oscillatory shearing commenced.

\begin{figure}[!t]
\includegraphics[width=1.0\columnwidth]{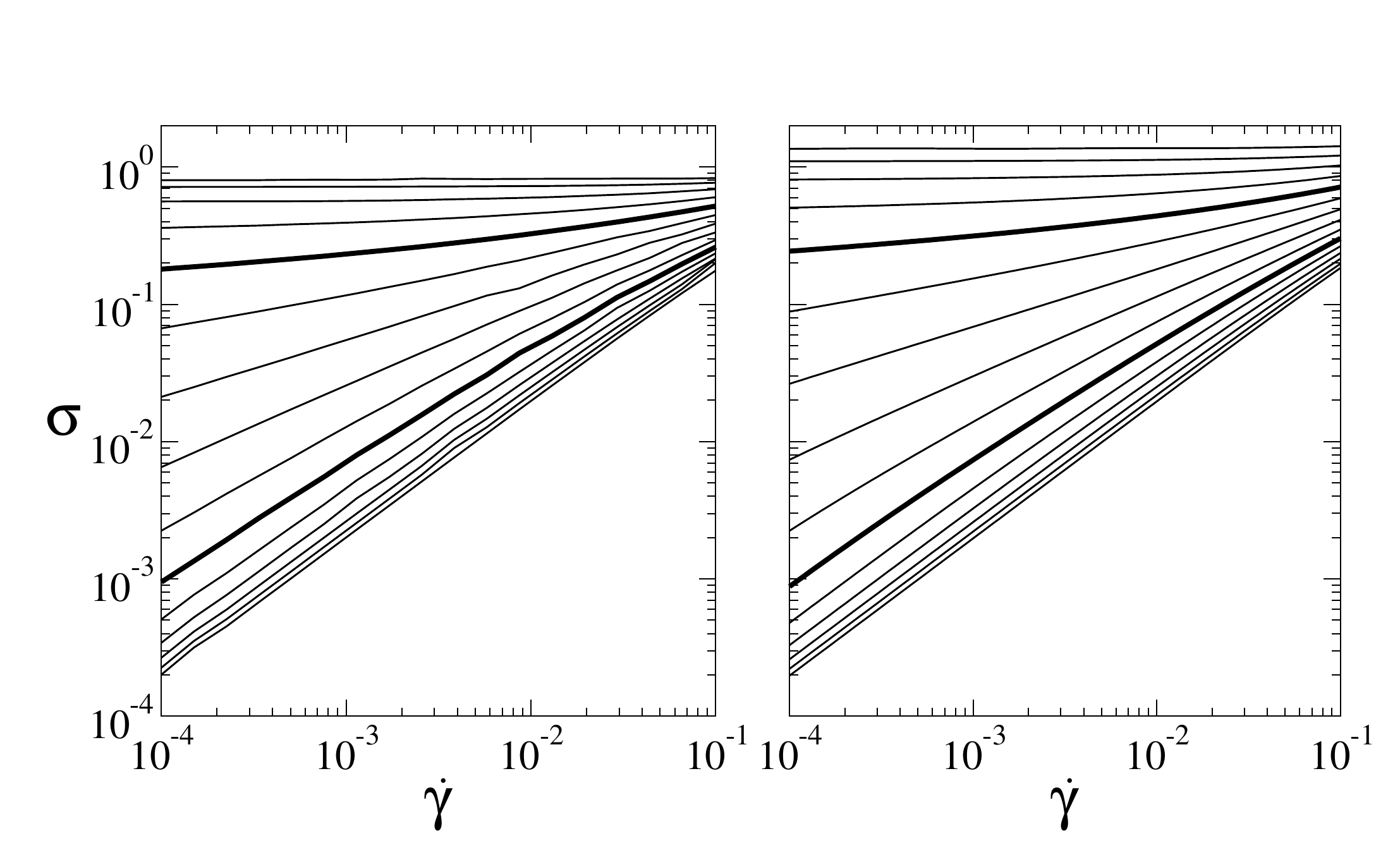}
\caption{Steady state flow curves. {\bf Left:} full SGR model, obtained by solving the integral constitutive equation of Ref.~\cite{ISI:000074893400095}. {\bf Right:} simplified SGR model. In both cases,  effective temperature values $x=0.2,0.4,0.6\cdots 3.0$ (curves downwards), with curves for $x=1.0$ and $x=2.0$ shown in bold. In the simplified SGR model, $E_{\rm max}=18.0$ and $N=32$.} 
\label{fig:flowCurves}
\end{figure}

Results for the real and imaginary parts of $G^*(\omega,t)$,
$G'(\omega,t)$ and $G''(\omega,t)$, are shown for the full SGR model
by the solid and dashed lines respectively in
Fig.~\ref{fig:viscospectra}. The left panel shows results for a fixed
sample age $t$ at a noise temperature in the glass phase. The right
panel shows results above the glass transition temperature. The
spectra show a broad power-law dependence on frequency, consistent
with the model's underlying spectrum of relaxation timescales.
Corresponding results for the simplified SGR model are shown by
symbols in the same figure. Excellent agreement with the full SGR
model again substantiates our claim that the simplified model exactly
agrees with the full model in the linear rheological regime.

\subsection{Nonlinear steady state flow curves}

Having discussed the linear rheological regime, in which the full and
simplified models exactly coincide, we now address nonlinear flows. We
start by considering the steady state relationship between the shear
stress and shear rate, as encoded in the flow curve,
$\sigma(\gdot)$. Results for this quantity computed in the full SGR
model are shown in the left panel of Fig.~\ref{fig:flowCurves}. For
high noise temperatures, $x > 2$, the model displays Newtonian flow
response in which $\sigma\sim
\gdot$. For intermediate noise temperatures, $1<x<2$, it shows power-law fluid behaviour in which $\sigma\sim \gdot^{x-1}$. For noise temperature in the glass phase, $x<1$, the  flow curve displays a yield stress $\sigmay(x)$, such that  $\sigma(\gdot)-\sigmay(x)\sim \gdot^{1-x}$. The yield stress $\sigmay$ shows a linear onset with $\xg-x$ below the glass point.

Corresponding results computed within the simplified SGR model are
shown in the right panel of the same figure. All the same quantitative
features as in the full SGR model are preserved, but with qualitative
differences of about a factor $2$ between the full and simplified
models in the most strongly linear flows, {\it i.e.}, in the glass
phase.

\subsection{Dynamical nonlinear rheology: imposed strain}

\subsubsection{Shear startup from rest}

\begin{figure}[!t]
\includegraphics[width=1.0\columnwidth]{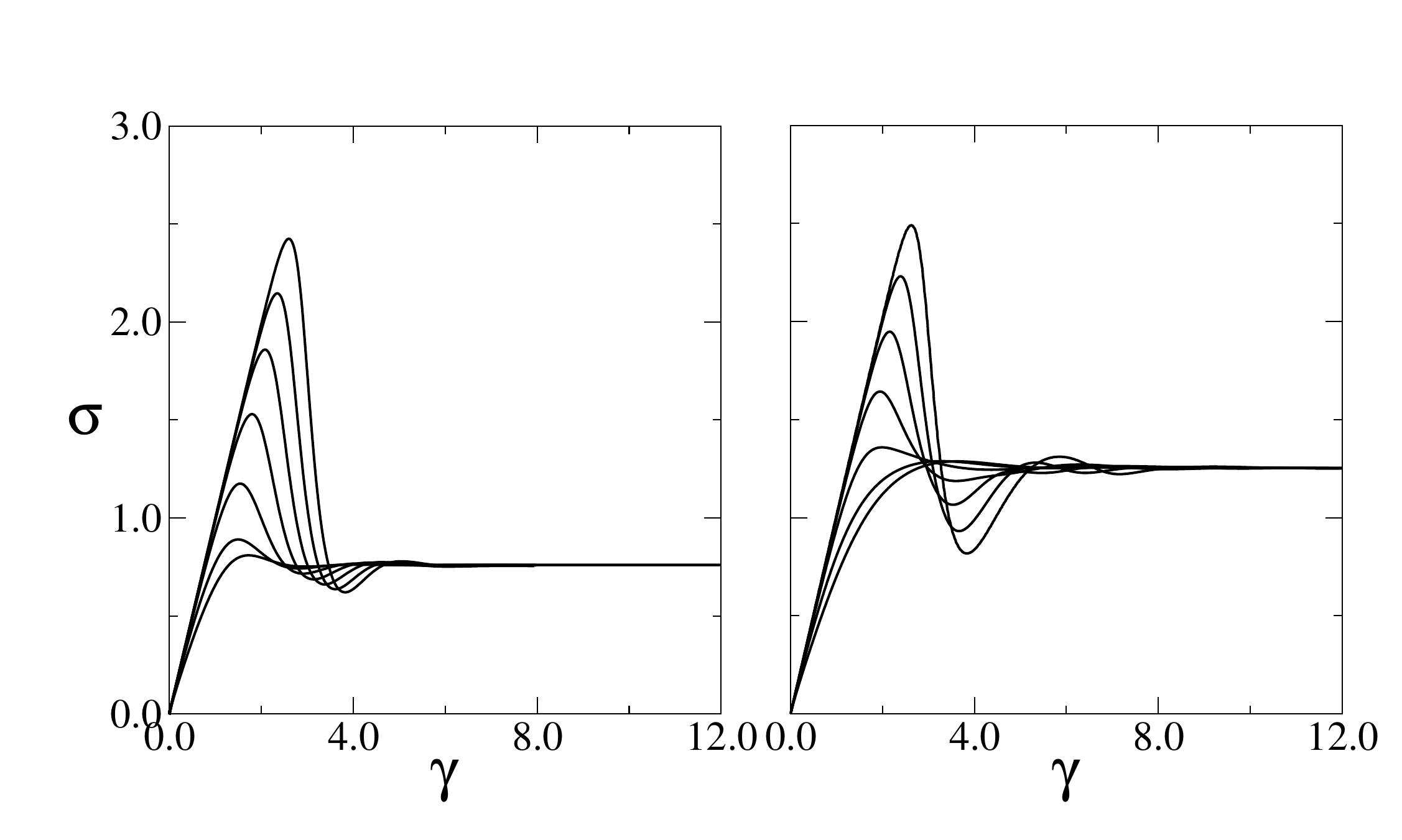}
\caption{Shear startup following the imposition of a step shear rate. {\bf Left:} full SGR model, obtained by solving the integral constitutive equation of Ref.~\cite{ISI:000074893400095}. {\bf Right:} simplified SGR model. In both cases, effective temperature $x=0.3$, imposed shear rate $\gdot_0=0.001$, initial sample ages $\tw=10^2,10^3,10^4,10^5,10^6,10^7,10^8$ (peak values upwards). In the simplified SGR model, $E_{\rm max}=12.0$, $N=32$, $dt=0.03$.} 
\label{fig:startup}
\end{figure}

Consider now a startup experiment in which a shear of rate $\gdot_0$
is suddenly switched on at time $\tw$, with the shear rate held
constant thereafter. We thus have $\gdot(t)=\gdot_0
\Theta(t-\tw)$, where $\Theta$ is the Heaviside function.

The left panel of Fig.~\ref{fig:startup} shows results for the stress
response as a function of accumulating strain,
$\gamma(t)=\gdot_0(t-\tw)$, computed in the glass phase of the full SGR
model. At early time intervals, for which the accumulated strain is
modest, the model shows an elastic solid-like response in which the
stress increases linearly with strain, $\sigma(t)=\gamma(t)$,
consistent with elements being in deep enough traps that their plastic
relaxation is initially negligible. By contrast, in the limit of long
times $t\to\infty$ and large strains $\gamma\to\infty$, the sample
flows in a liquid-like way, with the stress assuming a steady state
value prescribed by the flow curve $\sigma(\gdot_0)$ described in the
previous subsection. These early-time solid-like and late-time
liquid-like responses accordingly show no dependence on the age of the
sample before the shearing commenced. In contrast, at intermediate
strains the stress overshoots its final steady state value, and the
size of this overshoot shows a strong dependence on the sample age,
$\tw$.

The right panel of Fig.~\ref{fig:startup} shows corresponding results
computed within the simplified SGR model, for matched parameter
values. All the qualitative features are preserved in moving from the
full to simplified SGR model, and with only modest quantitative
differences.

So far, then, we have seen that the simplified SGR model exactly
reproduces the predictions of the full model in the regime of linear
rheology. We have further shown that it reproduces the full model's
qualitative behaviour in the nonlinear steady state flow curve, and in
nonlinear shear startup, with modest quantitative differences. Having
thus developed some confidence in the simplified SGR model, we now
proceed to present some new rheological predictions within the
simplified model that have not, to this author's knowledge, been
previously computed within full SGR.

\subsubsection{Shear rate jumps}

Having discussed in the previous subsection shear startup from an
initial rest state, $\gdot=0$, to a shear rate $\gdot_0$, we turn now to
consider a protocol in which a sample is sheared at some initial rate
$\gdot_1$ until it attains a steady flowing state, and is then subject
at some time $t=0$ to a shear rate jump to a final value $\gdot_2$. To
this author's knowledge, such a protocol has not previously been
studied in the full SGR model. All the results presented here are
computed within the simplified model.

\begin{figure}[!t]
\includegraphics[width=1.0\columnwidth]{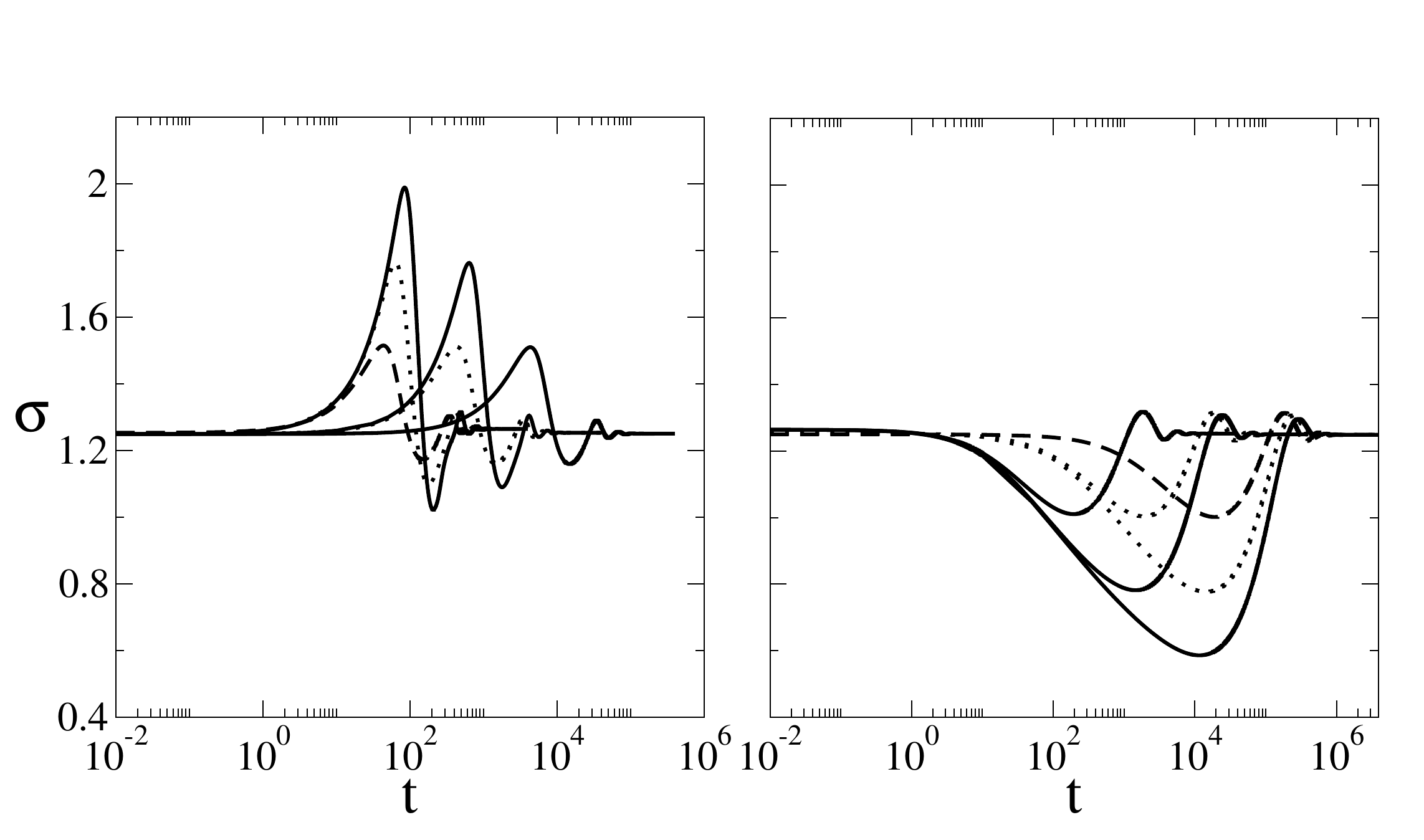}
\caption{Stress evolution following a shear rate jump in the simplified SGR model. In each run, the system is first evolved to steady state at an initial shear rate $\gdot_1$. Then, at a time defined to be $t=0$, the shear rate is jumped either up or down to $\gdot_2$ and the stress is plotted as a function of the time $t$ since that jump. {\bf Left:} upward strain rate jumps. Solid lines: $\gdot_1=10^{-5}$ with $\gdot_2=10^{-4},10^{-3},10^{-2}$. Dotted lines: $\gdot_1=10^{-4}$ with $\gdot_2=10^{-3},10^{-2}$. Dashed line: $\gdot_1=10^{-3}$ with $\gdot_2=10^{-2}$. At any fixed $\gdot_1$, times of stress peak move right with decreasing $\gdot_2$.  {\bf Right:} downward strain rate jumps. Solid lines: $\gdot_1=10^{-2}$ with $\gdot_2=10^{-3},10^{-4},10^{-5}$ Dotted lines: $\gdot_1=10^{-3}$ with $\gdot_2=10^{-4},10^{-5}$. Dashed line: $\gdot_1=10^{-4}$ with $\gdot_2=10^{-5}$. At any fixed $\gdot_1$, times of stress dip move rightward with decreasing $\gdot_2$.  $x=0.3$.  $dt =0.01,E_{\rm max}=12.0,N=32$.}
\label{fig:jumps}
\end{figure}

In the left panel of Fig.~\ref{fig:jumps}, we show results for the
time-dependent stress response $\sigma(t)$ in several different upward
strain rate jumps, $\gdot_2>\gdot_1$. In each case, the stress starts
at early times at its value as prescribed by the steady state flow
curve, $\sigma(\gdot_1)$, and tends at late times to a different value
that is also prescribed by the steady state flow curve,
$\sigma(\gdot_2)$. (Because the flow curve is a rather flat function of
shear rate for $\gdot\ll 1$ in the glass phase, $x<1$, these initial
and final values, $\sigma(\gdot_1)$ and $\sigma(\gdot_2)$, are rather
similar.)

Between these short- and long-time asymptotes, the stress displays an
overshoot that depends on both $\gdot_1$ and $\gdot_2$. The time at
which the overshoot occurs appears to scale roughly as $t\sim
\gdot_2^{-1}$, to within logarithmic corrections set by $\gdot_1$. The
stress overshoot accordingly happens when the accumulated strain
approaches a value $O(1)$ (to within logarithmic corrections),
consistent with elements then being pulled out of their traps by the
imposed strain. The height of the overshoot is set by
$\gdot_2/\gdot_1$.

The right panel of Fig.~\ref{fig:jumps} shows counterpart results for
downward strain rate jumps, $\gdot_2<\gdot_1$. In this case the stress
signal shows an undershoot in between its initial and final steady
state values. The time at which this undershoot occurs increases with
decreasing $\gdot_2$, and its height is set by $\gdot_2/\gdot_1$.

\vspace{0.4cm}
\subsubsection{Flow curve sweeps}

Consider now a protocol in which a sample is presheared to a steady
flowing state by executing $\gamma_{\rm preshear}$ strain units at a
high shear rate $\gdot_{\rm max}$. The strain rate is then stepped
downwards in $N_{\rm sweep}$ logarithmic increments to a low strain rate
$\gdot_{\rm min}$, waiting a time $\Delta t$ at each strain rate value
before further reducing the strain rate by a constant factor
$(\gdot_{\rm max}/\gdot_{\rm min})^{1/N_{\rm sweep}}$. Once $\gdot_{\rm min}$ is
attained the sweep is reversed, with the strain rate stepped upwards
through the same $N_{\rm sweep}$ values of strain rate, spending the same time
$\Delta t$ at each strain rate.

\begin{figure}[!t]
\includegraphics[width=1.0\columnwidth]{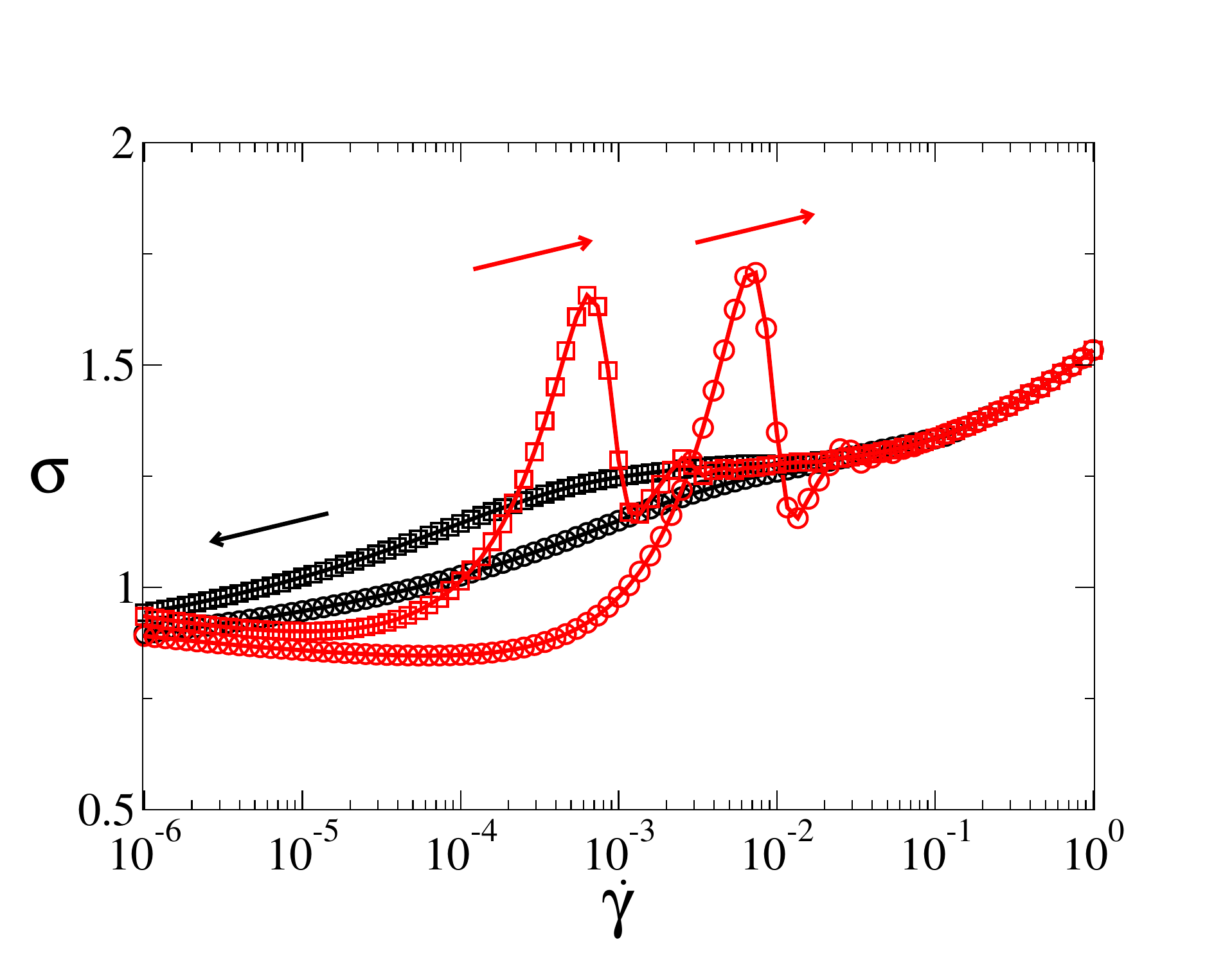}
\caption{Rheological hysteresis in flow curve sweeps in the simplified SGR model. Two separate sweeps are shown. In each, the system is first sheared to steady state by executing $\gamma_{\rm preshear}=250$ strain units at a high shear rate $\gdot_{\rm max}=1.0$. The strain rate is then stepped downwards to a low strain rate $\gdot_{\rm min}=10^{-6}$, waiting a time $\Delta t$ at each strain rate value before reducing the strain rate by a constant factor $(\gdot_{\rm max}/\gdot_{\rm min})^{1/N_{\rm sweep}}$. The strain rate is then stepped upwards through the same $N_{\rm sweep}$ values of strain rate with the same time $\Delta t$ spent at each shear rate. $N_{\rm sweep}=90$. Circles: $\Delta t=25.0$. Squares: $\Delta t=250.0$. Black symbols: down-sweep. Red symbols: up-sweep. Effective temperature $x=0.3$.  $dt =0.01,E_{\rm max}=12.0,N=32$. } 
\label{fig:sweep}
\end{figure}

Fig.~\ref{fig:sweep} shows results for the stress obtained by the
final time for each strain rate value, plotted as a function of that
strain rate, for two different values of $\Delta t$. In each case, the
black symbols denote the initial down-sweep and the red symbols the
later up-sweep.  These curves show all the same features as in the
full SGR model~\cite{ISI:000396030800011}, which can be summarised as
follows.

Notable hysteresis is clearly evident between the down- and up-sweeps.
Consider first the down-sweep. For strain rates higher than about
$10^{-2}$, the stress lies on the steady state flow curve, which has a
slight upward curvature as a function of $\gdot$. At lower strain
rates, the stress falls away from the steady state flow curve. This
occurs because the system cannot age into deeper traps quickly enough
to keep pace with the ever decreasing strain rate. Accordingly, the
sample remains in a more fluid-like state than it would be at a true
steady state for any imposed strain rate. The viscosity and shear
stress therefore remain low compared with the values they would assume
on the true steady state flow curve.

Once the strain rate reaches $\gdot_{\rm min}=10^{-6}$, the up-sweep
is commenced. During this upsweep, the stress initially ({\it i.e.,}
at low strain rates) lies below that seen during the
down-sweep. Indeed, it even decreases with increasing $\gdot$. This is
because the stress response of the SGR model is intrinsically
viscoelastic. Accordingly, the sample retains some memory of the
stress it had accumulated at the earlier high values of strain rate
during the down-sweep, which is still slowly relaxing even as the
shear rate increases again during the up-sweep.

This regime of declining stress ends with a steep upturn in the stress
as the strain rate increases yet further. The up-sweep stress then
rises above its down-sweep counterpart and indeed overshoots its
flow-curve value, before finally declining to meet the steady state
flow curve at the highest strain rates. This overshoot is the
counterpart to that seen in shear startup in Fig.~\ref{fig:startup},
and other upward shear rate jumps in Fig.~\ref{fig:jumps}. The shear
rate at which the overshoot occurs is seen to scale as $1/\Delta
t$. Indeed, all the features just described shift a decade to the left
between the curves for $\Delta t=25.0$ and $\Delta t =250.0$.

\subsubsection{Large amplitude oscillatory shear (LAOS)}

\begin{figure}[!t]
\includegraphics[width=1.0\columnwidth]{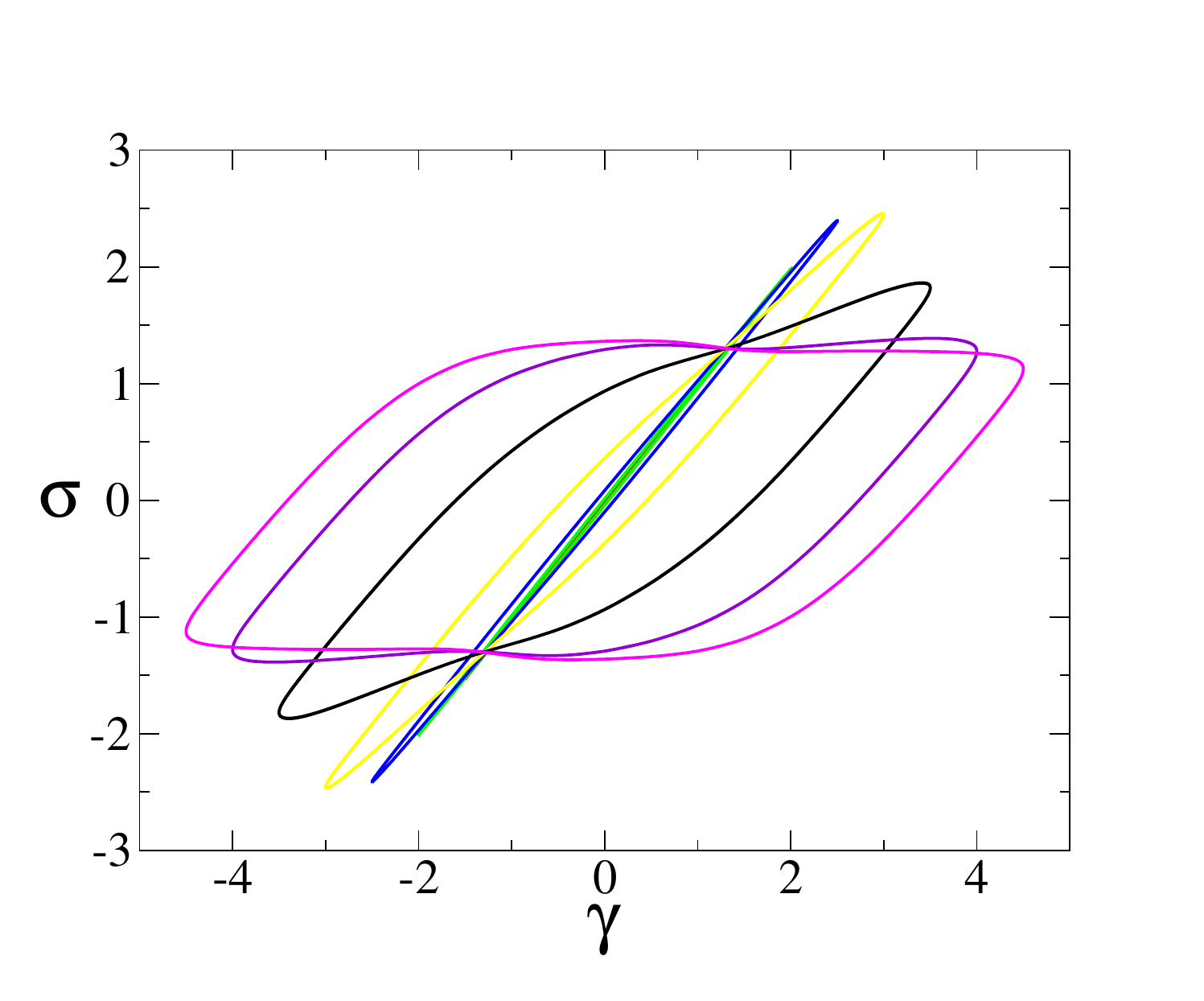}
\caption{Lissajous-Bowditch figures showing a parametric plot of stress $\sigma(t)$ against strain $\gamma(t)$ during large amplitude oscillatory shear (LAOS) in the  simplified SGR model.  Effective temperature $x=0.3$ and initial sample age $\tw=10$. The imposed strain $\gamma(t)=\gamma_0\sin(\omega t)$ with $\omega=0.01$ and $\gamma_0=1.0,1.5,2.0\cdots 4.5$  (curves outwards). For each value of $\gamma_0$, data are shown for the 99th and 100th cycles, with the data for these two cycles not being discernible from each other by eye.   $dt =0.005,E_{\rm max}=12.0,N=32$} 
\label{fig:lissajous}
\end{figure}

Let us consider now a protocol in which a sample is freshly prepared
at an initial time $t=0$, then left to age undisturbed to a time $\tw$
before an oscillatory shear strain is commenced,
$\gamma(t)=\gamma_0\sin(\omega t)$. We present here results obtained
within the simplified SGR model. LAOS has been previously studied in
the full SGR model (extended to allow shear banding) in
Refs.~\cite{ISI:000427032600013,ISI:000390226100005}.

For values of $\gamma_0$ in the nonlinear regime, we find that the
system attains after many strain cycles a state in which the stress
response is invariant under cycle-to-cycle translations, $t\to
t+2\pi/\omega$. For small $\gamma_0$, the sample instead continues to
age slightly from cycle to cycle, as seen in the viscoelastic spectra
of Fig.~\ref{fig:viscospectra}, left. Fig.~\ref{fig:lissajous} shows 
parametric so-called Lissajous-Bowditch (LB) plots of the stress
$\sigma(t)$ as a function of strain $\gamma(t)$ for the 99th and 100th
cycles, indeed with no discernible difference in the stress response
between these two cycles. Curves are shown for several values of the
imposed strain amplitude, $\gamma_0$, for a fixed frequency
$\omega$. The value of $\gamma_0$ in each case can be simply read off
from the maximum value of $\gamma(t)$ attained during the cycle.

For low values of $\gamma_0$, each LB curve takes the form of a highly
elongated, almost needle-like ellipse, oriented so as to have a slope
$\sigma'(\gamma)\approx 1$ over much of the cycle (except, obviously,
at the turning points of maximum and minimum strain). This is
consistent with the SGR model showing rather elastic behaviour at
stresses below the yield stress, with modulus $k=1$ in our units. For
the intermediate strain amplitudes explored, $\gamma_0=3.0$ and
$\gamma_0=3.5$, the LB curves instead adopt a characteristic diamond
shape, again with a slope $\approx 1$ for stresses below the yield
stress, but now with a reduced slope for higher stresses.  At the
highest strain amplitudes $\gamma_0$, the LB curves likewise show a
slope $\approx 1$ for stresses below the yield stress, but are much
flatter above yield.

The same progression in the shapes of the LB curves with increasing
strain amplitude $\gamma_0$ is also seen in the full SGR model.
Curves such as these have been discussed in detail in the LAOS
literature for yield stress fluids in terms of a sequence of physical
processes, with elastic caging at low stresses and yielding at higher
stresses~\cite{ISI:000287095800012}.

In Ref.~\cite{ISI:000312240900002}, experimental data for a carbopol
gel in large amplitude oscillatory shear (albeit in that work for
LAOStress rather than LAOStrain) was compared with the so-called
elastic Herschel-Bulkley (EHB) model. That model is constructed to
give a stress linear in strain below the yield stress, and a
Herschel-Bulkley relationship between stress and strain rate above
yield (as in the steady state flow curve of the SGR model). The EHB
model fails to capture the diamond shaped LB curves seen
experimentally in Ref.~\cite{ISI:000312240900002}, and reproduced here
for intermediate values of strain amplitudes in SGR. In particular,
the part of these diamond-shaped LB curves above yield shows a
hardening {\em compared with EHB}. (Care is warranted with
nomenclature here, because this part of the curve represents a {\em
softening} relative to the elastic regime below yield.)

Motivated by this observed hardening relative to EHB, many recent
attempts to build constitutive models of elastoplastic rheology have
incorporated the concept of `kinematic hardening'. This is typically
discussed as modelling the movement of the centre of a material's yield
surface, and is captured by including in the constitutive model
equations an additional variable termed the `back-stress'. Attempts to
justify this back-stress in terms of underlying mesoscopic physics
however remain largely unsatisfactory to date.

A pleasing feature of SGR is that it naturally captures these diamond
shaped LB curves (and many other features of elastoplastic rheology
besides) without recourse to the notion of a back-stress. How any
effective back-stress emerges from the SGR model remains an
interesting question. Feasibly, it could represent one of the next
higher moments of the local strain distribution, besides the average
strain encoded in the first moment.

\subsubsection{Strain cycling: Bauschinger effect}

\begin{figure}[!t]
\includegraphics[width=1.0\columnwidth]{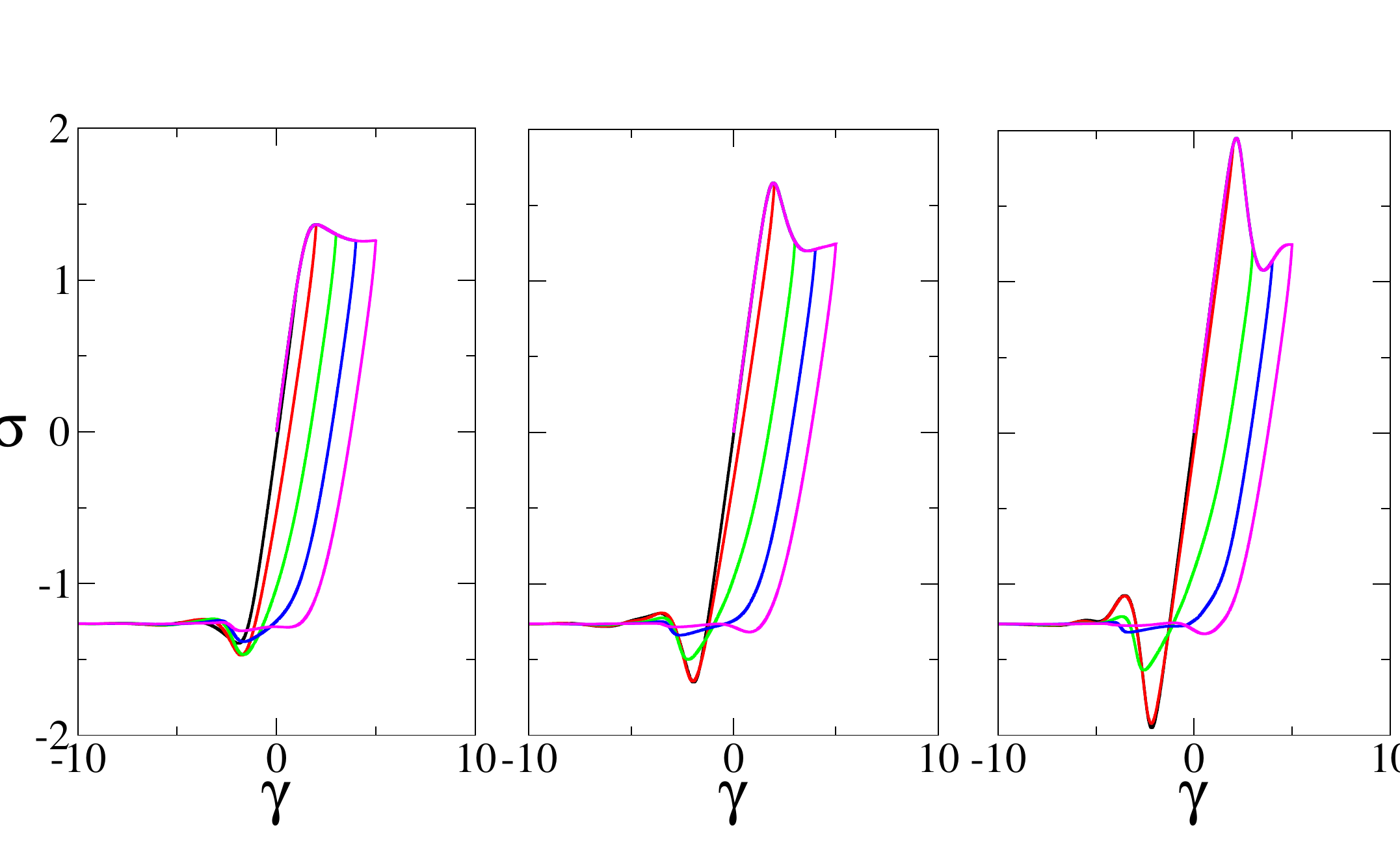}
\caption{Age-dependent Bauschinger effect in the simplified SGR model. Parametric plot  of stress $\sigma(t)$ against strain $\gamma(t)$ in (first) a forward strain by $\gamma_0$ strain units at a rate $\gdot=0.01$ followed (second) by a reverse strain at a rate $\gdot=-0.01$ to a final strain of $-10$ units.  Effective temperature $x=0.3$. Initial sample age before shearing commences: {\bf (left)} $\tw=10^3$, {\bf (middle)} $\tw=10^4$, {\bf (right)} $\tw=10^5$. In each panel, the imposed forward strain $\gamma_0=1.0, 2.0, 3.0, 4.0, 5.0$  in curves rightwards. $dt =0.01,E_{\rm max}=12.0,N=32$.} 
\label{fig:bauschinger}
\end{figure}

In 1886, Bauschinger reported an effective reduction in the tensile
yield stress of a polycrystalline metal following a tensile pre-strain
in the opposite direction~\cite{bauschinger1886veranderung}. The
effect has since also been discussed in the context of shear
deformations. We shall now explore this Bauschinger effect within the
simplified SGR model. To do so, let us consider a protocol in which a
sample is freshly prepared at time $t=0$, then left to age undisturbed
before a forward shear rate $\gdot$ is applied up to a forward strain
$\gamma_0$. The strain is then reversed at an equal and opposite
applied shear rate, $-\gdot$, up to a final strain of $-10$ units.

Figure~\ref{fig:bauschinger} shows the predictions of the simplified
SGR model in this protocol. The initial sample age $\tw$ increases by
a factor $10$ between each successive panel from left to right across
the figure. In each panel, results are shown for several values of the
total forward strain $\gamma_0$ applied before the strain direction is
reversed.

During the initial forward straining phase SGR predicts a stress
overshoot. Indeed, we have already discussed this in the context of a
simple forward shear startup experiment. Recall
Fig.~\ref{fig:startup}. The associated yield stress and strain
increase with increasing sample age $\tw$ across the panels of
Fig.~\ref{fig:bauschinger} from left to right. The degree to which
this stress overshoot and subsequent stress decline are explored, for
any $\tw$, increases with increasing $\gamma_0$. For values of
$\gamma_0$ large enough to give significant yielding in this forward
direction, a reduced yield stress is then observed during the
subsequent backward straining. This corresponds to a Bauschinger
effect (asymmetry between the initial forward and subsequent backward
yield stress), the size of which increases with increasing age of the
sample $\tw$ before the first, forward shear commenced.

\subsection{Dynamical nonlinear rheology: imposed stress}

%
%

\begin{figure}[!t]
\includegraphics[width=1.0\columnwidth]{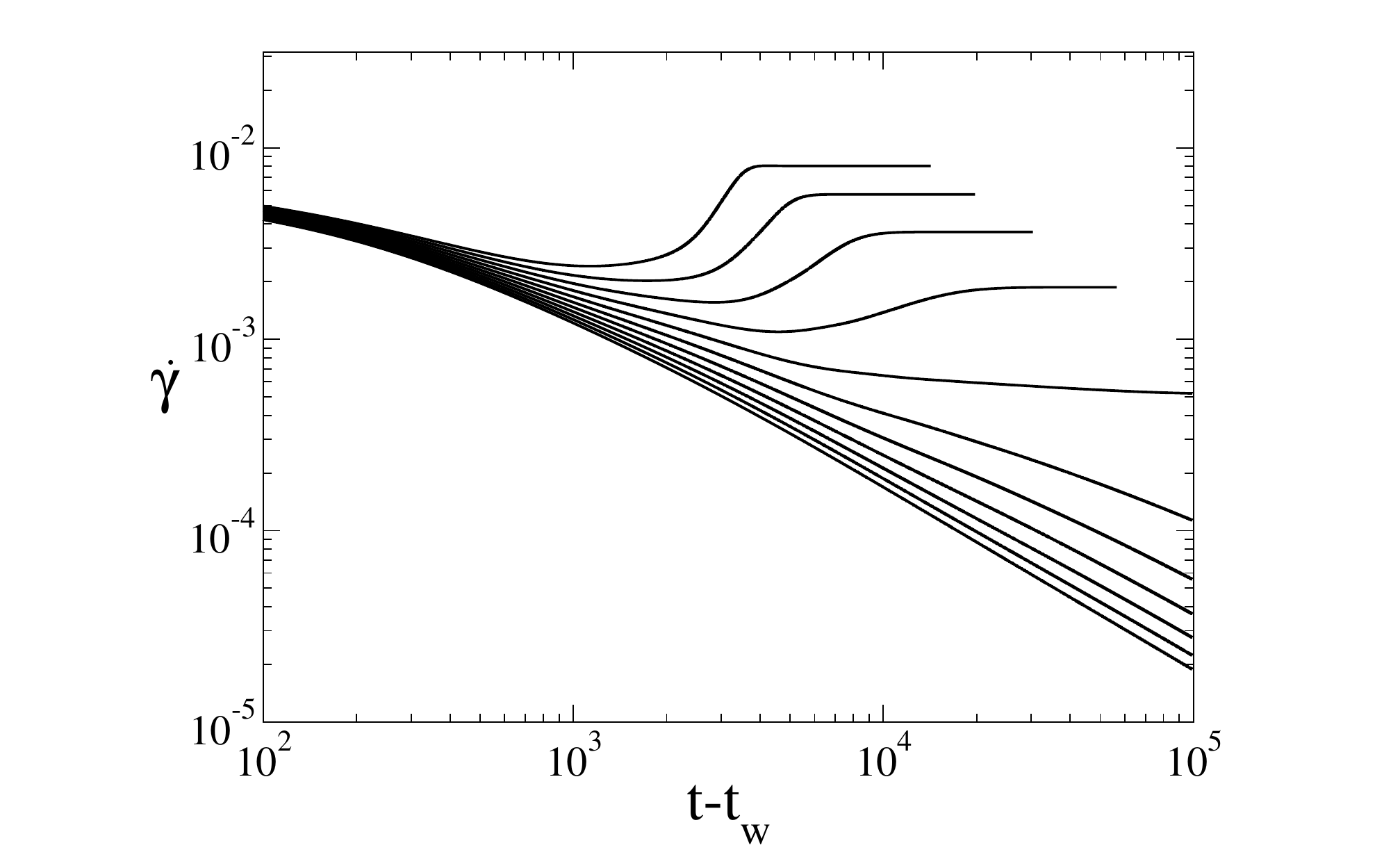}
\caption{Creep and (for imposed stress values $\sigma_0>\sigmay$) yielding following the imposition of a step stress of amplitude $\sigma_0$ at time $\tw$ in the simplified SGR model. Effective temperature $x=0.3$, initial sample age $\tw=10^3$. Imposed stress values, scaled by the yield stress, are $\sigma_0/\sigmay=0.990,0.992\cdots1.010$ (curves upwards).  $dt =0.01,E_{\rm max}=8.0,N=64$.} 
\label{fig:creep}
\end{figure}

So far, we have discussed the predictions of the simplified SGR model
for rheological protocols in which the shear strain is imposed as a
function of time. We turn finally to a common stress-imposed
protocol. In particular, we consider a sample that is freshly prepared
at time $t=0$ and left to age undisturbed up to a time $\tw$, when a
step shear stress of amplitude $\sigma_0$ is suddenly applied. The
imposed stress is accordingly $\sigma(t)=\sigma_0\Theta(t-\tw)$, where
$\Theta$ is the Heaviside function.

The strain rate response as a function of time $t-\tw$ is shown in
Fig.~\ref{fig:creep} for a fixed value of the sample age $\tw$ and
several values of the imposed stress $\sigma_0$ from below to above
the yield stress $\sigmay$ (defined as the stress attained in the
limit $\gdot\to 0$ of the steady state flow curve). 

For the smallest imposed stress values shown, the shear rate decays as a
function of time, 
and the
corresponding strain response increases sublinearly.
In this way, the material creeps ever forward, but
at an ever declining shear rate, never attaining a flowing state of
constant non-zero $\gdot$.
For the imposed stresses just above $\sigmay$, the sample initially
displays a window of sublinear creep in which the shear rate
progressively decreases, much as it would for an imposed stress below
yield. In marked contrast, however, at later times the sample yields
and the shear rate suddenly increases to attain its value as
prescribed by the steady state flow curve $\sigma(\gdot)$.  The same
behaviour was reported (in the format of strain versus time) in the
full SGR model in Ref.~\cite{ISI:000085655200008}.

This ability of the SGR model to capture power creep followed by
fluidisation and yielding following the imposition of a step shear
stress just above the yield stress should be particularly
noted. Reports of such behaviour in other constitutive models of
elastoplastic fluids (at least in those that have monotonic underlying
constitutive curves $\sigma(\gdot)$, precluding steady state shear
banding) are rare~\cite{ISI:000423131200025}: it appears difficult to
capture this complicated behaviour in a simple constitutive model
with just a small number of dynamical variables.

\section{Possible tensorial generalisation}
\label{sec:tensorial}
 
So far, we have presented a simplified SGR model with only a
scalarised shear stress. In order to perform CFD, one  needs a
constitutive model with a fully tensorial stress. We offer finally one
possible choice for such a model, following here the simplest path to
tensorialising the scalar model discussed above, and leaving other more
sophisticated generalisations for future work.
 
We define the stress carried in wells of depth $E$ as
 \be
\label{eqn:newLabel}
 \tens{\Sigma}(E,t)=G(E,t)\tens{l}(E,t),
 \ee
 where we have defined a new  tensorial strain variable $\tens{l}$. We then write the evolution equations
 \begin{widetext}
 \bea
 \left(\frac{d}{dt}+\vecv{v}\cdot\nabla\right)G(E,t)&=&-\frac{G(E,t)}{\tau(E,\tens{l}(E,t))}+\rho(E)Y(t),\nonumber\\
 \left(\frac{d}{dt}+\vecv{v}\cdot\nabla\right)\tens{\Sigma}(E,t)&=&\tens{\Sigma}(E,t)\cdot\tens{K}+\tens{K}^T\cdot\tens{\Sigma}(E,t)+2G(E,t)\tens{D}-\frac{\tens{\Sigma}(E,t)}{\tau(E,\tens{l}(E,t))}.
 \eea
 \end{widetext}
 Here $\tens{K}$ is the velocity gradient tensor, and $\tens{D}$ its symmetric part.
 
 We then have the usual definition of the hopping rate:
 \be
 Y(t)=\int dE\frac{G(E,t)}{\tau(E,\tens{l}(E,t))}
 \ee
and the prior distribution
 \be
 \rho(E)=\exp(-E).
 \ee 
The time constant is now defined as
 \be
 \tau(E,l)=\tau_0 \exp\left(\frac{E-I(\tens{l})}{x}\right),
 \ee
in which $I$ is a suitable invariant of the local strain tensor
$\tens{l}$, for which we suggest $I=\tfrac{1}{2}\tens{l}:\tens{l}$. We note that the strain tensor, $\tens{l}$, is defined by the ratio of the stress tensor, $\tens{\Sigma}$, and the modulus, $G$, via Eqn.~\ref{eqn:newLabel}.

To compute the response of this tensorial model in homogeneous flow
would require the evolution of $7N$ time-differential equations: one
at each $N$ for $G$ and $6$ for each independent component of the
symmetric tensor $\tens{\Sigma}$, having again discretised $E$ on a
grid of $N$ values. To perform CFD would require $7N$ such variables
at each lattice site, with three additional variables for the flow
velocity vector.

 
\section{Conclusions}
\label{sec:conclusions}

In this work, we have introduced a simplified constitutive model for
the elastoviscoplastic rheology of yield stress fluids, motivated by
the widely used soft glassy rheology model. We have demonstrated this
simplified model to capture a wide array of observed rheological
behaviours, in both strain-imposed and stress-imposed flow protocols,
in both the linear and nonlinear rheological regimes. Once suitable
units of modulus, length and time are chosen, the model has only
one dimensionless parameter: the effective noise temperature, $x$.

The original soft glassy rheology model on which this simplified model
is based has been widely used in the literature. However, the
computation within it of even homogeneous simple shear flows is
considerably cumbersome, involving the solution of a partial
differential equation, $\partial_tP(E,l,t)=\cdots$, or the solution of
two coupled nonlinear integral equations, or the direct simulation of
typically $10^5$ hopping SGR elements. 

In contrast, the computation of homogeneous simple shear flows within
the simplified model requires the time-evolution of only $2N$
relatively simple differential equations, with values of $N$ as low as
$16$ giving good results. This renders SGR much more readily
accessible to anyone wishing to fit its predictions to rheometric
data. And whereas the original model was prohibitively costly for use
in CFD to address flows in complicated geometries, or complicated
flows arising due to spontaneous symmetry breaking instabilities even
in simple geometries, the simplified model is now sufficiently simple
for use in CFD, once suitably tensorialised. Indeed, work is currently
in progress to benchmark its behaviour in the canonical CFD geometries
of 2D flow past a cylinder and 3D flow past a sphere.

{\it Acknowledgements ---} The author thanks Andrew Clarke for discussions, and Mike Cates, Gareth McKinley and Ron Larson for feedback on the manuscript. Schlumberger Cambridge Research is acknowledged for funding.


\begin{thebibliography}{114}%
\makeatletter
\providecommand \@ifxundefined [1]{%
 \@ifx{#1\undefined}
}%
\providecommand \@ifnum [1]{%
 \ifnum #1\expandafter \@firstoftwo
 \else \expandafter \@secondoftwo
 \fi
}%
\providecommand \@ifx [1]{%
 \ifx #1\expandafter \@firstoftwo
 \else \expandafter \@secondoftwo
 \fi
}%
\providecommand \natexlab [1]{#1}%
\providecommand \enquote  [1]{``#1''}%
\providecommand \bibnamefont  [1]{#1}%
\providecommand \bibfnamefont [1]{#1}%
\providecommand \citenamefont [1]{#1}%
\providecommand \href@noop [0]{\@secondoftwo}%
\providecommand \href [0]{\begingroup \@sanitize@url \@href}%
\providecommand \@href[1]{\@@startlink{#1}\@@href}%
\providecommand \@@href[1]{\endgroup#1\@@endlink}%
\providecommand \@sanitize@url [0]{\catcode `\\12\catcode `\$12\catcode
  `\&12\catcode `\#12\catcode `\^12\catcode `\_12\catcode `\%12\relax}%
\providecommand \@@startlink[1]{}%
\providecommand \@@endlink[0]{}%
\providecommand \doibase [0]{http://dx.doi.org/}%
\providecommand \selectlanguage [0]{\@gobble}%
\providecommand \bibinfo  [0]{\@secondoftwo}%
\providecommand \bibfield  [0]{\@secondoftwo}%
\providecommand \translation [1]{[#1]}%
\providecommand \BibitemOpen [0]{}%
\providecommand \bibitemStop [0]{}%
\providecommand \bibitemNoStop [0]{.\EOS\space}%
\providecommand \EOS [0]{\spacefactor3000\relax}%
\providecommand \BibitemShut  [1]{\csname bibitem#1\endcsname}%
\let\auto@bib@innerbib\@empty
\bibitem [{\citenamefont {Bonn}\ \emph {et~al.}(2017)\citenamefont {Bonn},
  \citenamefont {Denn}, \citenamefont {Berthier}, \citenamefont {Divoux},\ and\
  \citenamefont {Manneville}}]{ISI:000407999000001}%
  \BibitemOpen
  \bibfield  {author} {\bibinfo {author} {\bibfnamefont {D.}~\bibnamefont
  {Bonn}}, \bibinfo {author} {\bibfnamefont {M.~M.}\ \bibnamefont {Denn}},
  \bibinfo {author} {\bibfnamefont {L.}~\bibnamefont {Berthier}}, \bibinfo
  {author} {\bibfnamefont {T.}~\bibnamefont {Divoux}}, \ and\ \bibinfo {author}
  {\bibfnamefont {S.}~\bibnamefont {Manneville}},\ }\href {\doibase
  {10.1103/RevModPhys.89.035005}} {\bibfield  {journal} {\bibinfo  {journal}
  {{Rev. Mod. Phys.}}\ }\textbf {\bibinfo {volume} {{89}}},\ \bibinfo {pages}
  {035005} (\bibinfo {year} {{2017}})}\BibitemShut {NoStop}%
\bibitem [{\citenamefont {Coussot}(2018)}]{ISI:000419993800001}%
  \BibitemOpen
  \bibfield  {author} {\bibinfo {author} {\bibfnamefont {P.}~\bibnamefont
  {Coussot}},\ }\href {\doibase {10.1007/s00397-017-1055-7}} {\bibfield
  {journal} {\bibinfo  {journal} {{Rheol. Acta}}\ }\textbf {\bibinfo {volume}
  {{57}}},\ \bibinfo {pages} {1} (\bibinfo {year} {{2018}})}\BibitemShut
  {NoStop}%
\bibitem [{\citenamefont {Coussot}(2014)}]{coussot2014yield}%
  \BibitemOpen
  \bibfield  {author} {\bibinfo {author} {\bibfnamefont {P.}~\bibnamefont
  {Coussot}},\ }\href@noop {} {\bibfield  {journal} {\bibinfo  {journal} {J.
  Non-Newton. Fluid}\ }\textbf {\bibinfo {volume} {211}},\ \bibinfo {pages}
  {31} (\bibinfo {year} {2014})}\BibitemShut {NoStop}%
\bibitem [{\citenamefont {Bonn}\ and\ \citenamefont
  {Denn}(2009)}]{ISI:000266878700032}%
  \BibitemOpen
  \bibfield  {author} {\bibinfo {author} {\bibfnamefont {D.}~\bibnamefont
  {Bonn}}\ and\ \bibinfo {author} {\bibfnamefont {M.~M.}\ \bibnamefont
  {Denn}},\ }\href {\doibase {10.1126/science.1174217}} {\bibfield  {journal}
  {\bibinfo  {journal} {{Science}}\ }\textbf {\bibinfo {volume} {{324}}},\
  \bibinfo {pages} {1401} (\bibinfo {year} {{2009}})}\BibitemShut {NoStop}%
\bibitem [{\citenamefont {Coussot}(2007)}]{ISI:000246368400004}%
  \BibitemOpen
  \bibfield  {author} {\bibinfo {author} {\bibfnamefont {P.}~\bibnamefont
  {Coussot}},\ }\href {\doibase {10.1039/b611021p}} {\bibfield  {journal}
  {\bibinfo  {journal} {{Soft Matter}}\ }\textbf {\bibinfo {volume} {{3}}},\
  \bibinfo {pages} {528} (\bibinfo {year} {{2007}})}\BibitemShut {NoStop}%
\bibitem [{\citenamefont {Denn}\ and\ \citenamefont
  {Bonn}(2011)}]{ISI:000293295000002}%
  \BibitemOpen
  \bibfield  {author} {\bibinfo {author} {\bibfnamefont {M.~M.}\ \bibnamefont
  {Denn}}\ and\ \bibinfo {author} {\bibfnamefont {D.}~\bibnamefont {Bonn}},\
  }\href {\doibase {10.1007/s00397-010-0504-3}} {\bibfield  {journal} {\bibinfo
   {journal} {{Rheol. Acta}}\ }\textbf {\bibinfo {volume} {{50}}},\ \bibinfo
  {pages} {307} (\bibinfo {year} {{2011}})}\BibitemShut {NoStop}%
\bibitem [{\citenamefont {Balmforth}\ \emph {et~al.}(2014)\citenamefont
  {Balmforth}, \citenamefont {Frigaard},\ and\ \citenamefont
  {Ovarlez}}]{balmforth2014yielding}%
  \BibitemOpen
  \bibfield  {author} {\bibinfo {author} {\bibfnamefont {N.~J.}\ \bibnamefont
  {Balmforth}}, \bibinfo {author} {\bibfnamefont {I.~A.}\ \bibnamefont
  {Frigaard}}, \ and\ \bibinfo {author} {\bibfnamefont {G.}~\bibnamefont
  {Ovarlez}},\ }\href@noop {} {\bibfield  {journal} {\bibinfo  {journal}
  {Annual Review of Fluid Mechanics}\ }\textbf {\bibinfo {volume} {46}},\
  \bibinfo {pages} {121} (\bibinfo {year} {2014})}\BibitemShut {NoStop}%
\bibitem [{\citenamefont {Moller}\ \emph {et~al.}(2006)\citenamefont {Moller},
  \citenamefont {Mewis},\ and\ \citenamefont {Bonn}}]{moller2006yield}%
  \BibitemOpen
  \bibfield  {author} {\bibinfo {author} {\bibfnamefont {P.~C.}\ \bibnamefont
  {Moller}}, \bibinfo {author} {\bibfnamefont {J.}~\bibnamefont {Mewis}}, \
  and\ \bibinfo {author} {\bibfnamefont {D.}~\bibnamefont {Bonn}},\ }\href@noop
  {} {\bibfield  {journal} {\bibinfo  {journal} {Soft matter}\ }\textbf
  {\bibinfo {volume} {2}},\ \bibinfo {pages} {274} (\bibinfo {year}
  {2006})}\BibitemShut {NoStop}%
\bibitem [{\citenamefont {Herschel}\ and\ \citenamefont
  {Bulkley}(1926)}]{ISI:000202025500001}%
  \BibitemOpen
  \bibfield  {author} {\bibinfo {author} {\bibfnamefont {W.}~\bibnamefont
  {Herschel}}\ and\ \bibinfo {author} {\bibfnamefont {R.}~\bibnamefont
  {Bulkley}},\ }\href {\doibase 10.1007/BF01432034} {\bibfield  {journal}
  {\bibinfo  {journal} {Kolloid-Zeitschrift}\ }\textbf {\bibinfo {volume}
  {39}},\ \bibinfo {pages} {291} (\bibinfo {year} {1926})}\BibitemShut
  {NoStop}%
\bibitem [{\citenamefont {Bingham}(1922)}]{bingham1922fluidity}%
  \BibitemOpen
  \bibfield  {author} {\bibinfo {author} {\bibfnamefont {E.~C.}\ \bibnamefont
  {Bingham}},\ }\href@noop {} {\emph {\bibinfo {title} {Fluidity and
  plasticity}}},\ Vol.~\bibinfo {volume} {2}\ (\bibinfo  {publisher}
  {McGraw-Hill},\ \bibinfo {year} {1922})\BibitemShut {NoStop}%
\bibitem [{\citenamefont {Larson}\ and\ \citenamefont
  {Wei}(2019)}]{larson2019review}%
  \BibitemOpen
  \bibfield  {author} {\bibinfo {author} {\bibfnamefont {R.~G.}\ \bibnamefont
  {Larson}}\ and\ \bibinfo {author} {\bibfnamefont {Y.}~\bibnamefont {Wei}},\
  }\href@noop {} {\bibfield  {journal} {\bibinfo  {journal} {J. Rheol.}\
  }\textbf {\bibinfo {volume} {63}},\ \bibinfo {pages} {477} (\bibinfo {year}
  {2019})}\BibitemShut {NoStop}%
\bibitem [{\citenamefont {Ikeda}\ \emph {et~al.}(2012)\citenamefont {Ikeda},
  \citenamefont {Berthier},\ and\ \citenamefont
  {Sollich}}]{ISI:000306323400007}%
  \BibitemOpen
  \bibfield  {author} {\bibinfo {author} {\bibfnamefont {A.}~\bibnamefont
  {Ikeda}}, \bibinfo {author} {\bibfnamefont {L.}~\bibnamefont {Berthier}}, \
  and\ \bibinfo {author} {\bibfnamefont {P.}~\bibnamefont {Sollich}},\ }\href
  {\doibase {10.1103/PhysRevLett.109.018301}} {\bibfield  {journal} {\bibinfo
  {journal} {{Phys. Rev. Lett.}}\ }\textbf {\bibinfo {volume} {{109}}},\
  \bibinfo {pages} {018301} (\bibinfo {year} {{2012}})}\BibitemShut {NoStop}%
\bibitem [{\citenamefont {Ikeda}\ \emph {et~al.}(2013)\citenamefont {Ikeda},
  \citenamefont {Berthier},\ and\ \citenamefont
  {Sollich}}]{ISI:000322230300001}%
  \BibitemOpen
  \bibfield  {author} {\bibinfo {author} {\bibfnamefont {A.}~\bibnamefont
  {Ikeda}}, \bibinfo {author} {\bibfnamefont {L.}~\bibnamefont {Berthier}}, \
  and\ \bibinfo {author} {\bibfnamefont {P.}~\bibnamefont {Sollich}},\ }\href
  {\doibase {10.1039/c3sm50503k}} {\bibfield  {journal} {\bibinfo  {journal}
  {{Soft Matter}}\ }\textbf {\bibinfo {volume} {{9}}},\ \bibinfo {pages} {7669}
  (\bibinfo {year} {{2013}})}\BibitemShut {NoStop}%
\bibitem [{\citenamefont {Sollich}\ \emph {et~al.}(1997)\citenamefont
  {Sollich}, \citenamefont {Lequeux}, \citenamefont {Hebraud},\ and\
  \citenamefont {Cates}}]{ISI:A1997WM06400048}%
  \BibitemOpen
  \bibfield  {author} {\bibinfo {author} {\bibfnamefont {P.}~\bibnamefont
  {Sollich}}, \bibinfo {author} {\bibfnamefont {F.}~\bibnamefont {Lequeux}},
  \bibinfo {author} {\bibfnamefont {P.}~\bibnamefont {Hebraud}}, \ and\
  \bibinfo {author} {\bibfnamefont {M.}~\bibnamefont {Cates}},\ }\href
  {\doibase {10.1103/PhysRevLett.78.2020}} {\bibfield  {journal} {\bibinfo
  {journal} {{Phys. Rev. Lett.}}\ }\textbf {\bibinfo {volume} {{78}}},\
  \bibinfo {pages} {2020} (\bibinfo {year} {{1997}})}\BibitemShut {NoStop}%
\bibitem [{\citenamefont {Fielding}\ \emph {et~al.}(2000)\citenamefont
  {Fielding}, \citenamefont {Sollich},\ and\ \citenamefont
  {Cates}}]{ISI:000085655200008}%
  \BibitemOpen
  \bibfield  {author} {\bibinfo {author} {\bibfnamefont {S.}~\bibnamefont
  {Fielding}}, \bibinfo {author} {\bibfnamefont {P.}~\bibnamefont {Sollich}}, \
  and\ \bibinfo {author} {\bibfnamefont {M.}~\bibnamefont {Cates}},\ }\href
  {\doibase {10.1122/1.551088}} {\bibfield  {journal} {\bibinfo  {journal} {{J.
  Rheol.}}\ }\textbf {\bibinfo {volume} {{44}}},\ \bibinfo {pages} {323}
  (\bibinfo {year} {{2000}})}\BibitemShut {NoStop}%
\bibitem [{\citenamefont {Sollich}(1998)}]{ISI:000074893400095}%
  \BibitemOpen
  \bibfield  {author} {\bibinfo {author} {\bibfnamefont {P.}~\bibnamefont
  {Sollich}},\ }\href {\doibase {10.1103/PhysRevE.58.738}} {\bibfield
  {journal} {\bibinfo  {journal} {{Phys. Rev. E}}\ }\textbf {\bibinfo {volume}
  {{58}}},\ \bibinfo {pages} {738} (\bibinfo {year} {{1998}})}\BibitemShut
  {NoStop}%
\bibitem [{\citenamefont {Mackley}\ \emph {et~al.}(1994)\citenamefont
  {Mackley}, \citenamefont {Marshall}, \citenamefont {Smeulders},\ and\
  \citenamefont {Zhao}}]{mackley1994rheological}%
  \BibitemOpen
  \bibfield  {author} {\bibinfo {author} {\bibfnamefont {M.}~\bibnamefont
  {Mackley}}, \bibinfo {author} {\bibfnamefont {R.}~\bibnamefont {Marshall}},
  \bibinfo {author} {\bibfnamefont {J.}~\bibnamefont {Smeulders}}, \ and\
  \bibinfo {author} {\bibfnamefont {F.}~\bibnamefont {Zhao}},\ }\href@noop {}
  {\bibfield  {journal} {\bibinfo  {journal} {Chemical Engineering Science}\
  }\textbf {\bibinfo {volume} {49}},\ \bibinfo {pages} {2551} (\bibinfo {year}
  {1994})}\BibitemShut {NoStop}%
\bibitem [{\citenamefont {Ketz}\ \emph {et~al.}(1988)\citenamefont {Ketz},
  \citenamefont {Prud'homme},\ and\ \citenamefont
  {Graessley}}]{ketz1988rheology}%
  \BibitemOpen
  \bibfield  {author} {\bibinfo {author} {\bibfnamefont {R.}~\bibnamefont
  {Ketz}}, \bibinfo {author} {\bibfnamefont {R.~K.}\ \bibnamefont
  {Prud'homme}}, \ and\ \bibinfo {author} {\bibfnamefont {W.}~\bibnamefont
  {Graessley}},\ }\href@noop {} {\bibfield  {journal} {\bibinfo  {journal}
  {Rheologica acta}\ }\textbf {\bibinfo {volume} {27}},\ \bibinfo {pages} {531}
  (\bibinfo {year} {1988})}\BibitemShut {NoStop}%
\bibitem [{\citenamefont {Khan}\ \emph {et~al.}(2000)\citenamefont {Khan},
  \citenamefont {Schnepper},\ and\ \citenamefont {Armstrong}}]{khan2000foam}%
  \BibitemOpen
  \bibfield  {author} {\bibinfo {author} {\bibfnamefont {S.~A.}\ \bibnamefont
  {Khan}}, \bibinfo {author} {\bibfnamefont {C.~A.}\ \bibnamefont {Schnepper}},
  \ and\ \bibinfo {author} {\bibfnamefont {R.~C.}\ \bibnamefont {Armstrong}},\
  }\href@noop {} {\bibfield  {journal} {\bibinfo  {journal} {Journal of
  Rheology}\ }\textbf {\bibinfo {volume} {32}},\ \bibinfo {pages} {69}
  (\bibinfo {year} {2000})}\BibitemShut {NoStop}%
\bibitem [{\citenamefont {Mason}\ \emph {et~al.}(1995)\citenamefont {Mason},
  \citenamefont {Bibette},\ and\ \citenamefont {Weitz}}]{mason1995elasticity}%
  \BibitemOpen
  \bibfield  {author} {\bibinfo {author} {\bibfnamefont {T.}~\bibnamefont
  {Mason}}, \bibinfo {author} {\bibfnamefont {J.}~\bibnamefont {Bibette}}, \
  and\ \bibinfo {author} {\bibfnamefont {D.}~\bibnamefont {Weitz}},\
  }\href@noop {} {\bibfield  {journal} {\bibinfo  {journal} {Physical review
  letters}\ }\textbf {\bibinfo {volume} {75}},\ \bibinfo {pages} {2051}
  (\bibinfo {year} {1995})}\BibitemShut {NoStop}%
\bibitem [{\citenamefont {Panizza}\ \emph {et~al.}(1996)\citenamefont
  {Panizza}, \citenamefont {Roux}, \citenamefont {Vuillaume}, \citenamefont
  {Lu},\ and\ \citenamefont {Cates}}]{panizza1996viscoelasticity}%
  \BibitemOpen
  \bibfield  {author} {\bibinfo {author} {\bibfnamefont {P.}~\bibnamefont
  {Panizza}}, \bibinfo {author} {\bibfnamefont {D.}~\bibnamefont {Roux}},
  \bibinfo {author} {\bibfnamefont {V.}~\bibnamefont {Vuillaume}}, \bibinfo
  {author} {\bibfnamefont {C.-Y.}\ \bibnamefont {Lu}}, \ and\ \bibinfo {author}
  {\bibfnamefont {M.}~\bibnamefont {Cates}},\ }\href@noop {} {\bibfield
  {journal} {\bibinfo  {journal} {Langmuir}\ }\textbf {\bibinfo {volume}
  {12}},\ \bibinfo {pages} {248} (\bibinfo {year} {1996})}\BibitemShut
  {NoStop}%
\bibitem [{\citenamefont {Hoffmann}\ and\ \citenamefont
  {Rauscher}(1993)}]{hoffmann1993aggregating}%
  \BibitemOpen
  \bibfield  {author} {\bibinfo {author} {\bibfnamefont {H.}~\bibnamefont
  {Hoffmann}}\ and\ \bibinfo {author} {\bibfnamefont {A.}~\bibnamefont
  {Rauscher}},\ }\href@noop {} {\bibfield  {journal} {\bibinfo  {journal}
  {Colloid and Polymer Science}\ }\textbf {\bibinfo {volume} {271}},\ \bibinfo
  {pages} {390} (\bibinfo {year} {1993})}\BibitemShut {NoStop}%
\bibitem [{\citenamefont {Mason}\ and\ \citenamefont
  {Weitz}(1995)}]{mason1995linear}%
  \BibitemOpen
  \bibfield  {author} {\bibinfo {author} {\bibfnamefont {T.}~\bibnamefont
  {Mason}}\ and\ \bibinfo {author} {\bibfnamefont {D.}~\bibnamefont {Weitz}},\
  }\href@noop {} {\bibfield  {journal} {\bibinfo  {journal} {Physical review
  letters}\ }\textbf {\bibinfo {volume} {75}},\ \bibinfo {pages} {2770}
  (\bibinfo {year} {1995})}\BibitemShut {NoStop}%
\bibitem [{\citenamefont {Ramos}\ and\ \citenamefont
  {Cipelletti}(2001)}]{ISI:000172642100029}%
  \BibitemOpen
  \bibfield  {author} {\bibinfo {author} {\bibfnamefont {L.}~\bibnamefont
  {Ramos}}\ and\ \bibinfo {author} {\bibfnamefont {L.}~\bibnamefont
  {Cipelletti}},\ }\href {\doibase {10.1103/PhysRevLett.87.245503}} {\bibfield
  {journal} {\bibinfo  {journal} {{Phys. Rev. Lett.}}\ }\textbf {\bibinfo
  {volume} {{87}}},\ \bibinfo {pages} {245503} (\bibinfo {year}
  {{2001}})}\BibitemShut {NoStop}%
\bibitem [{\citenamefont {Batista}\ \emph {et~al.}(2006)\citenamefont
  {Batista}, \citenamefont {Raymundo}, \citenamefont {Sousa}, \citenamefont
  {Empis},\ and\ \citenamefont {Franco}}]{batista2006colored}%
  \BibitemOpen
  \bibfield  {author} {\bibinfo {author} {\bibfnamefont {A.~P.}\ \bibnamefont
  {Batista}}, \bibinfo {author} {\bibfnamefont {A.}~\bibnamefont {Raymundo}},
  \bibinfo {author} {\bibfnamefont {I.}~\bibnamefont {Sousa}}, \bibinfo
  {author} {\bibfnamefont {J.}~\bibnamefont {Empis}}, \ and\ \bibinfo {author}
  {\bibfnamefont {J.~M.}\ \bibnamefont {Franco}},\ }\href@noop {} {\bibfield
  {journal} {\bibinfo  {journal} {Food Biophysics}\ }\textbf {\bibinfo {volume}
  {1}},\ \bibinfo {pages} {216} (\bibinfo {year} {2006})}\BibitemShut {NoStop}%
\bibitem [{\citenamefont {Papenhuijzen}(1972)}]{papenhuijzen1972role}%
  \BibitemOpen
  \bibfield  {author} {\bibinfo {author} {\bibfnamefont {J.}~\bibnamefont
  {Papenhuijzen}},\ }\href@noop {} {\bibfield  {journal} {\bibinfo  {journal}
  {Rheologica Acta}\ }\textbf {\bibinfo {volume} {11}},\ \bibinfo {pages} {73}
  (\bibinfo {year} {1972})}\BibitemShut {NoStop}%
\bibitem [{\citenamefont {Divoux}\ \emph {et~al.}(2010)\citenamefont {Divoux},
  \citenamefont {Tamarii}, \citenamefont {Barentin},\ and\ \citenamefont
  {Manneville}}]{divoux2010}%
  \BibitemOpen
  \bibfield  {author} {\bibinfo {author} {\bibfnamefont {T.}~\bibnamefont
  {Divoux}}, \bibinfo {author} {\bibfnamefont {D.}~\bibnamefont {Tamarii}},
  \bibinfo {author} {\bibfnamefont {C.}~\bibnamefont {Barentin}}, \ and\
  \bibinfo {author} {\bibfnamefont {S.}~\bibnamefont {Manneville}},\
  }\href@noop {} {\bibfield  {journal} {\bibinfo  {journal} {Physical Review
  Letters}\ }\textbf {\bibinfo {volume} {104}},\ \bibinfo {pages} {208301}
  (\bibinfo {year} {2010})}\BibitemShut {NoStop}%
\bibitem [{\citenamefont {Divoux}\ \emph
  {et~al.}(2011{\natexlab{a}})\citenamefont {Divoux}, \citenamefont
  {Barentin},\ and\ \citenamefont {Manneville}}]{ISI:000295085700080}%
  \BibitemOpen
  \bibfield  {author} {\bibinfo {author} {\bibfnamefont {T.}~\bibnamefont
  {Divoux}}, \bibinfo {author} {\bibfnamefont {C.}~\bibnamefont {Barentin}}, \
  and\ \bibinfo {author} {\bibfnamefont {S.}~\bibnamefont {Manneville}},\
  }\href {\doibase {10.1039/c1sm05740e}} {\bibfield  {journal} {\bibinfo
  {journal} {{Soft Matter}}\ }\textbf {\bibinfo {volume} {{7}}},\ \bibinfo
  {pages} {9335} (\bibinfo {year} {{2011}}{\natexlab{a}})}\BibitemShut
  {NoStop}%
\bibitem [{\citenamefont {Martin}\ and\ \citenamefont
  {T.~Hu}(2012)}]{Martinetal2012a}%
  \BibitemOpen
  \bibfield  {author} {\bibinfo {author} {\bibfnamefont {J.~D.}\ \bibnamefont
  {Martin}}\ and\ \bibinfo {author} {\bibfnamefont {Y.}~\bibnamefont {T.~Hu}},\
  }\href {\doibase 10.1039/C2SM25299F} {\bibfield  {journal} {\bibinfo
  {journal} {Soft Matter}\ }\textbf {\bibinfo {volume} {8}},\ \bibinfo {pages}
  {6940} (\bibinfo {year} {2012})}\BibitemShut {NoStop}%
\bibitem [{\citenamefont {Gibaud}\ \emph {et~al.}(2008)\citenamefont {Gibaud},
  \citenamefont {Barentin},\ and\ \citenamefont {Manneville}}]{gibaud2008}%
  \BibitemOpen
  \bibfield  {author} {\bibinfo {author} {\bibfnamefont {T.}~\bibnamefont
  {Gibaud}}, \bibinfo {author} {\bibfnamefont {C.}~\bibnamefont {Barentin}}, \
  and\ \bibinfo {author} {\bibfnamefont {S.}~\bibnamefont {Manneville}},\
  }\href@noop {} {\bibfield  {journal} {\bibinfo  {journal} {Physical Review
  Letters}\ }\textbf {\bibinfo {volume} {101}},\ \bibinfo {pages} {258302}
  (\bibinfo {year} {2008})}\BibitemShut {NoStop}%
\bibitem [{\citenamefont {Kurokawa}\ \emph {et~al.}(2015)\citenamefont
  {Kurokawa}, \citenamefont {Vidal}, \citenamefont {Kurita}, \citenamefont
  {Divoux},\ and\ \citenamefont {Manneville}}]{Kurokawa}%
  \BibitemOpen
  \bibfield  {author} {\bibinfo {author} {\bibfnamefont {A.}~\bibnamefont
  {Kurokawa}}, \bibinfo {author} {\bibfnamefont {V.}~\bibnamefont {Vidal}},
  \bibinfo {author} {\bibfnamefont {K.}~\bibnamefont {Kurita}}, \bibinfo
  {author} {\bibfnamefont {T.}~\bibnamefont {Divoux}}, \ and\ \bibinfo {author}
  {\bibfnamefont {S.}~\bibnamefont {Manneville}},\ }\href@noop {} {\bibfield
  {journal} {\bibinfo  {journal} {Soft Matter}\ }\textbf {\bibinfo {volume}
  {11}},\ \bibinfo {pages} {9026} (\bibinfo {year} {2015})}\BibitemShut
  {NoStop}%
\bibitem [{\citenamefont {Koumakis}\ and\ \citenamefont
  {Petekidis}(2011)}]{koumakis2011two}%
  \BibitemOpen
  \bibfield  {author} {\bibinfo {author} {\bibfnamefont {N.}~\bibnamefont
  {Koumakis}}\ and\ \bibinfo {author} {\bibfnamefont {G.}~\bibnamefont
  {Petekidis}},\ }\href@noop {} {\bibfield  {journal} {\bibinfo  {journal}
  {Soft Matter}\ }\textbf {\bibinfo {volume} {7}},\ \bibinfo {pages} {2456}
  (\bibinfo {year} {2011})}\BibitemShut {NoStop}%
\bibitem [{\citenamefont {Liddel}\ and\ \citenamefont
  {Boger}(1996)}]{liddel1996yield}%
  \BibitemOpen
  \bibfield  {author} {\bibinfo {author} {\bibfnamefont {P.~V.}\ \bibnamefont
  {Liddel}}\ and\ \bibinfo {author} {\bibfnamefont {D.~V.}\ \bibnamefont
  {Boger}},\ }\href@noop {} {\bibfield  {journal} {\bibinfo  {journal} {Journal
  of non-newtonian fluid mechanics}\ }\textbf {\bibinfo {volume} {63}},\
  \bibinfo {pages} {235} (\bibinfo {year} {1996})}\BibitemShut {NoStop}%
\bibitem [{\citenamefont {Dimitriou}\ and\ \citenamefont
  {McKinley}(2014)}]{ISI:000341025500004}%
  \BibitemOpen
  \bibfield  {author} {\bibinfo {author} {\bibfnamefont {C.~J.}\ \bibnamefont
  {Dimitriou}}\ and\ \bibinfo {author} {\bibfnamefont {G.~H.}\ \bibnamefont
  {McKinley}},\ }\href {\doibase {10.1039/c4sm00578c}} {\bibfield  {journal}
  {\bibinfo  {journal} {{Soft Matter}}\ }\textbf {\bibinfo {volume} {{10}}},\
  \bibinfo {pages} {6619} (\bibinfo {year} {{2014}})}\BibitemShut {NoStop}%
\bibitem [{\citenamefont {Divoux}\ \emph
  {et~al.}(2011{\natexlab{b}})\citenamefont {Divoux}, \citenamefont
  {Barentin},\ and\ \citenamefont {Manneville}}]{ISI:000294447600069}%
  \BibitemOpen
  \bibfield  {author} {\bibinfo {author} {\bibfnamefont {T.}~\bibnamefont
  {Divoux}}, \bibinfo {author} {\bibfnamefont {C.}~\bibnamefont {Barentin}}, \
  and\ \bibinfo {author} {\bibfnamefont {S.}~\bibnamefont {Manneville}},\
  }\href {\doibase {10.1039/c1sm05607g}} {\bibfield  {journal} {\bibinfo
  {journal} {{Soft Matter}}\ }\textbf {\bibinfo {volume} {{7}}},\ \bibinfo
  {pages} {8409} (\bibinfo {year} {{2011}}{\natexlab{b}})}\BibitemShut
  {NoStop}%
\bibitem [{\citenamefont {Magnin}\ and\ \citenamefont
  {Piau}(1990)}]{ISI:A1990DZ50400005}%
  \BibitemOpen
  \bibfield  {author} {\bibinfo {author} {\bibfnamefont {A.}~\bibnamefont
  {Magnin}}\ and\ \bibinfo {author} {\bibfnamefont {J.~M.}\ \bibnamefont
  {Piau}},\ }\href {\doibase {10.1016/0377-0257(90)85005-J}} {\bibfield
  {journal} {\bibinfo  {journal} {{J. Nonnewton. Fluid Mech.}}\ }\textbf
  {\bibinfo {volume} {{36}}},\ \bibinfo {pages} {85} (\bibinfo {year}
  {{1990}})}\BibitemShut {NoStop}%
\bibitem [{\citenamefont {Gibaud}\ \emph {et~al.}(2010)\citenamefont {Gibaud},
  \citenamefont {Frelat},\ and\ \citenamefont
  {Manneville}}]{ISI:000280140800011}%
  \BibitemOpen
  \bibfield  {author} {\bibinfo {author} {\bibfnamefont {T.}~\bibnamefont
  {Gibaud}}, \bibinfo {author} {\bibfnamefont {D.}~\bibnamefont {Frelat}}, \
  and\ \bibinfo {author} {\bibfnamefont {S.}~\bibnamefont {Manneville}},\
  }\href {\doibase {10.1039/c000886a}} {\bibfield  {journal} {\bibinfo
  {journal} {{Soft Matter}}\ }\textbf {\bibinfo {volume} {{6}}},\ \bibinfo
  {pages} {3482} (\bibinfo {year} {{2010}})}\BibitemShut {NoStop}%
\bibitem [{\citenamefont {Grenard}\ \emph {et~al.}(2014)\citenamefont
  {Grenard}, \citenamefont {Divoux}, \citenamefont {Taberlet},\ and\
  \citenamefont {Manneville}}]{ISI:000332461800012}%
  \BibitemOpen
  \bibfield  {author} {\bibinfo {author} {\bibfnamefont {V.}~\bibnamefont
  {Grenard}}, \bibinfo {author} {\bibfnamefont {T.}~\bibnamefont {Divoux}},
  \bibinfo {author} {\bibfnamefont {N.}~\bibnamefont {Taberlet}}, \ and\
  \bibinfo {author} {\bibfnamefont {S.}~\bibnamefont {Manneville}},\ }\href
  {\doibase {10.1039/c3sm52548a}} {\bibfield  {journal} {\bibinfo  {journal}
  {{Soft Matter}}\ }\textbf {\bibinfo {volume} {{10}}},\ \bibinfo {pages}
  {1555} (\bibinfo {year} {{2014}})}\BibitemShut {NoStop}%
\bibitem [{\citenamefont {Sprakel}\ \emph {et~al.}(2011)\citenamefont
  {Sprakel}, \citenamefont {Lindstr{\"o}m}, \citenamefont {Kodger},\ and\
  \citenamefont {Weitz}}]{sprakel2011stress}%
  \BibitemOpen
  \bibfield  {author} {\bibinfo {author} {\bibfnamefont {J.}~\bibnamefont
  {Sprakel}}, \bibinfo {author} {\bibfnamefont {S.~B.}\ \bibnamefont
  {Lindstr{\"o}m}}, \bibinfo {author} {\bibfnamefont {T.~E.}\ \bibnamefont
  {Kodger}}, \ and\ \bibinfo {author} {\bibfnamefont {D.~A.}\ \bibnamefont
  {Weitz}},\ }\href@noop {} {\bibfield  {journal} {\bibinfo  {journal}
  {Physical review letters}\ }\textbf {\bibinfo {volume} {106}},\ \bibinfo
  {pages} {248303} (\bibinfo {year} {2011})}\BibitemShut {NoStop}%
\bibitem [{\citenamefont {Bauer}\ \emph {et~al.}(2006)\citenamefont {Bauer},
  \citenamefont {Oberdisse},\ and\ \citenamefont
  {Ramos}}]{bauer2006collective}%
  \BibitemOpen
  \bibfield  {author} {\bibinfo {author} {\bibfnamefont {T.}~\bibnamefont
  {Bauer}}, \bibinfo {author} {\bibfnamefont {J.}~\bibnamefont {Oberdisse}}, \
  and\ \bibinfo {author} {\bibfnamefont {L.}~\bibnamefont {Ramos}},\
  }\href@noop {} {\bibfield  {journal} {\bibinfo  {journal} {Physical review
  letters}\ }\textbf {\bibinfo {volume} {97}},\ \bibinfo {pages} {258303}
  (\bibinfo {year} {2006})}\BibitemShut {NoStop}%
\bibitem [{\citenamefont {Sentjabrskaja}\ \emph {et~al.}(2015)\citenamefont
  {Sentjabrskaja}, \citenamefont {Chaudhuri}, \citenamefont {Hermes},
  \citenamefont {Poon}, \citenamefont {Horbach}, \citenamefont {Egelhaaf},\
  and\ \citenamefont {Laurati}}]{ISI:000357577400001}%
  \BibitemOpen
  \bibfield  {author} {\bibinfo {author} {\bibfnamefont {T.}~\bibnamefont
  {Sentjabrskaja}}, \bibinfo {author} {\bibfnamefont {P.}~\bibnamefont
  {Chaudhuri}}, \bibinfo {author} {\bibfnamefont {M.}~\bibnamefont {Hermes}},
  \bibinfo {author} {\bibfnamefont {W.~C.~K.}\ \bibnamefont {Poon}}, \bibinfo
  {author} {\bibfnamefont {J.}~\bibnamefont {Horbach}}, \bibinfo {author}
  {\bibfnamefont {S.~U.}\ \bibnamefont {Egelhaaf}}, \ and\ \bibinfo {author}
  {\bibfnamefont {M.}~\bibnamefont {Laurati}},\ }\href {\doibase
  {10.1038/srep11884}} {\bibfield  {journal} {\bibinfo  {journal} {{Sci.
  Rep.}}\ }\textbf {\bibinfo {volume} {{5}}},\ \bibinfo {pages} {11884}
  (\bibinfo {year} {{2015}})}\BibitemShut {NoStop}%
\bibitem [{\citenamefont {Siebenb{\"u}rger}\ \emph {et~al.}(2012)\citenamefont
  {Siebenb{\"u}rger}, \citenamefont {Ballauff},\ and\ \citenamefont
  {Voigtmann}}]{siebenburger2012creep}%
  \BibitemOpen
  \bibfield  {author} {\bibinfo {author} {\bibfnamefont {M.}~\bibnamefont
  {Siebenb{\"u}rger}}, \bibinfo {author} {\bibfnamefont {M.}~\bibnamefont
  {Ballauff}}, \ and\ \bibinfo {author} {\bibfnamefont {T.}~\bibnamefont
  {Voigtmann}},\ }\href@noop {} {\bibfield  {journal} {\bibinfo  {journal}
  {Physical review letters}\ }\textbf {\bibinfo {volume} {108}},\ \bibinfo
  {pages} {255701} (\bibinfo {year} {2012})}\BibitemShut {NoStop}%
\bibitem [{\citenamefont {Yoshimura}\ and\ \citenamefont
  {Prud'homme}(1987)}]{Yoshimura1987}%
  \BibitemOpen
  \bibfield  {author} {\bibinfo {author} {\bibfnamefont {A.~S.}\ \bibnamefont
  {Yoshimura}}\ and\ \bibinfo {author} {\bibfnamefont {R.~K.}\ \bibnamefont
  {Prud'homme}},\ }\href {\doibase 10.1007/BF01333843} {\bibfield  {journal}
  {\bibinfo  {journal} {Rheol. Acta}\ }\textbf {\bibinfo {volume} {26}},\
  \bibinfo {pages} {428} (\bibinfo {year} {1987})}\BibitemShut {NoStop}%
\bibitem [{\citenamefont {Knaebel}\ \emph {et~al.}(2000)\citenamefont
  {Knaebel}, \citenamefont {Bellour}, \citenamefont {Munch}, \citenamefont
  {Viasnoff}, \citenamefont {Lequeux},\ and\ \citenamefont
  {Harden}}]{Knaebel2000}%
  \BibitemOpen
  \bibfield  {author} {\bibinfo {author} {\bibfnamefont {A.}~\bibnamefont
  {Knaebel}}, \bibinfo {author} {\bibfnamefont {M.}~\bibnamefont {Bellour}},
  \bibinfo {author} {\bibfnamefont {J.-P.}\ \bibnamefont {Munch}}, \bibinfo
  {author} {\bibfnamefont {V.}~\bibnamefont {Viasnoff}}, \bibinfo {author}
  {\bibfnamefont {F.}~\bibnamefont {Lequeux}}, \ and\ \bibinfo {author}
  {\bibfnamefont {J.~L.}\ \bibnamefont {Harden}},\ }\href
  {http://stacks.iop.org/0295-5075/52/i=1/a=073} {\bibfield  {journal}
  {\bibinfo  {journal} {Europhys. Lett.}\ }\textbf {\bibinfo {volume} {52}},\
  \bibinfo {pages} {73} (\bibinfo {year} {2000})}\BibitemShut {NoStop}%
\bibitem [{\citenamefont {Viasnoff}\ \emph {et~al.}(2003)\citenamefont
  {Viasnoff}, \citenamefont {Jurine},\ and\ \citenamefont
  {Lequeux}}]{Viasnoff2003}%
  \BibitemOpen
  \bibfield  {author} {\bibinfo {author} {\bibfnamefont {V.}~\bibnamefont
  {Viasnoff}}, \bibinfo {author} {\bibfnamefont {S.}~\bibnamefont {Jurine}}, \
  and\ \bibinfo {author} {\bibfnamefont {F.}~\bibnamefont {Lequeux}},\ }\href
  {\doibase 10.1039/b204377g} {\bibfield  {journal} {\bibinfo  {journal}
  {Faraday Discuss.}\ }\textbf {\bibinfo {volume} {123}},\ \bibinfo {pages}
  {253} (\bibinfo {year} {2003})}\BibitemShut {NoStop}%
\bibitem [{\citenamefont {Rouyer}\ \emph {et~al.}(2008)\citenamefont {Rouyer},
  \citenamefont {Cohen-Addad}, \citenamefont {H{\"{o}}hler}, \citenamefont
  {Sollich},\ and\ \citenamefont {Fielding}}]{Rouyer2008}%
  \BibitemOpen
  \bibfield  {author} {\bibinfo {author} {\bibfnamefont {F.}~\bibnamefont
  {Rouyer}}, \bibinfo {author} {\bibfnamefont {S.}~\bibnamefont {Cohen-Addad}},
  \bibinfo {author} {\bibfnamefont {R.}~\bibnamefont {H{\"{o}}hler}}, \bibinfo
  {author} {\bibfnamefont {P.}~\bibnamefont {Sollich}}, \ and\ \bibinfo
  {author} {\bibfnamefont {S.~M.}\ \bibnamefont {Fielding}},\ }\href {\doibase
  10.1140/epje/i2008-10382-7} {\bibfield  {journal} {\bibinfo  {journal} {Eur.
  Phys. J. E}\ }\textbf {\bibinfo {volume} {27}},\ \bibinfo {pages} {309}
  (\bibinfo {year} {2008})}\BibitemShut {NoStop}%
\bibitem [{\citenamefont {Ewoldt}\ \emph {et~al.}(2010)\citenamefont {Ewoldt},
  \citenamefont {Winter}, \citenamefont {Maxey},\ and\ \citenamefont
  {McKinley}}]{ISI:000273854300007}%
  \BibitemOpen
  \bibfield  {author} {\bibinfo {author} {\bibfnamefont {R.~H.}\ \bibnamefont
  {Ewoldt}}, \bibinfo {author} {\bibfnamefont {P.}~\bibnamefont {Winter}},
  \bibinfo {author} {\bibfnamefont {J.}~\bibnamefont {Maxey}}, \ and\ \bibinfo
  {author} {\bibfnamefont {G.~H.}\ \bibnamefont {McKinley}},\ }\href {\doibase
  {10.1007/s00397-009-0403-7}} {\bibfield  {journal} {\bibinfo  {journal}
  {{Rheol. Acta}}\ }\textbf {\bibinfo {volume} {{49}}},\ \bibinfo {pages} {191}
  (\bibinfo {year} {{2010}})}\BibitemShut {NoStop}%
\bibitem [{\citenamefont {Renou}\ \emph {et~al.}(2010)\citenamefont {Renou},
  \citenamefont {Stellbrink},\ and\ \citenamefont {Petekidis}}]{Renou2010}%
  \BibitemOpen
  \bibfield  {author} {\bibinfo {author} {\bibfnamefont {F.}~\bibnamefont
  {Renou}}, \bibinfo {author} {\bibfnamefont {J.}~\bibnamefont {Stellbrink}}, \
  and\ \bibinfo {author} {\bibfnamefont {G.}~\bibnamefont {Petekidis}},\ }\href
  {\doibase {10.1122/1.3483610}} {\bibfield  {journal} {\bibinfo  {journal} {J.
  Rheol.}\ }\textbf {\bibinfo {volume} {54}},\ \bibinfo {pages} {1219}
  (\bibinfo {year} {2010})}\BibitemShut {NoStop}%
\bibitem [{\citenamefont {van~der Vaart}\ \emph {et~al.}(2013)\citenamefont
  {van~der Vaart}, \citenamefont {Rahmani}, \citenamefont {Zargar},
  \citenamefont {Hu}, \citenamefont {Bonn},\ and\ \citenamefont
  {Schall}}]{VanderVaart2013}%
  \BibitemOpen
  \bibfield  {author} {\bibinfo {author} {\bibfnamefont {K.}~\bibnamefont
  {van~der Vaart}}, \bibinfo {author} {\bibfnamefont {Y.}~\bibnamefont
  {Rahmani}}, \bibinfo {author} {\bibfnamefont {R.}~\bibnamefont {Zargar}},
  \bibinfo {author} {\bibfnamefont {Z.}~\bibnamefont {Hu}}, \bibinfo {author}
  {\bibfnamefont {D.}~\bibnamefont {Bonn}}, \ and\ \bibinfo {author}
  {\bibfnamefont {P.}~\bibnamefont {Schall}},\ }\href {\doibase
  10.1122/1.4808054} {\bibfield  {journal} {\bibinfo  {journal} {J. Rheol.}\
  }\textbf {\bibinfo {volume} {57}},\ \bibinfo {pages} {1195} (\bibinfo {year}
  {2013})}\BibitemShut {NoStop}%
\bibitem [{\citenamefont {Koumakis}\ \emph {et~al.}(2013)\citenamefont
  {Koumakis}, \citenamefont {Brady},\ and\ \citenamefont
  {Petekidis}}]{Koumakis2013}%
  \BibitemOpen
  \bibfield  {author} {\bibinfo {author} {\bibfnamefont {N.}~\bibnamefont
  {Koumakis}}, \bibinfo {author} {\bibfnamefont {J.~F.}\ \bibnamefont {Brady}},
  \ and\ \bibinfo {author} {\bibfnamefont {G.}~\bibnamefont {Petekidis}},\
  }\href@noop {} {\bibfield  {journal} {\bibinfo  {journal} {Phys. Rev. Lett.}\
  }\textbf {\bibinfo {volume} {110}},\ \bibinfo {pages} {178301} (\bibinfo
  {year} {2013})}\BibitemShut {NoStop}%
\bibitem [{\citenamefont {Poulos}\ \emph {et~al.}(2013)\citenamefont {Poulos},
  \citenamefont {Stellbrink},\ and\ \citenamefont {Petekidis}}]{Poulos2013}%
  \BibitemOpen
  \bibfield  {author} {\bibinfo {author} {\bibfnamefont {A.~S.}\ \bibnamefont
  {Poulos}}, \bibinfo {author} {\bibfnamefont {J.}~\bibnamefont {Stellbrink}},
  \ and\ \bibinfo {author} {\bibfnamefont {G.}~\bibnamefont {Petekidis}},\
  }\href {\doibase {10.1007/s00397-013-0703-9}} {\bibfield  {journal} {\bibinfo
   {journal} {Rheol. Acta}\ }\textbf {\bibinfo {volume} {52}},\ \bibinfo
  {pages} {785} (\bibinfo {year} {2013})}\BibitemShut {NoStop}%
\bibitem [{\citenamefont {Poulos}\ \emph {et~al.}(2015)\citenamefont {Poulos},
  \citenamefont {Renou}, \citenamefont {Jacob}, \citenamefont {Koumakis},\ and\
  \citenamefont {Petekidis}}]{Poulos2015}%
  \BibitemOpen
  \bibfield  {author} {\bibinfo {author} {\bibfnamefont {A.~S.}\ \bibnamefont
  {Poulos}}, \bibinfo {author} {\bibfnamefont {F.}~\bibnamefont {Renou}},
  \bibinfo {author} {\bibfnamefont {A.~R.}\ \bibnamefont {Jacob}}, \bibinfo
  {author} {\bibfnamefont {N.}~\bibnamefont {Koumakis}}, \ and\ \bibinfo
  {author} {\bibfnamefont {G.}~\bibnamefont {Petekidis}},\ }\href {\doibase
  {10.1007/s00397-015-0865-8}} {\bibfield  {journal} {\bibinfo  {journal}
  {Rheol. Acta}\ }\textbf {\bibinfo {volume} {54}},\ \bibinfo {pages} {715}
  (\bibinfo {year} {2015})}\BibitemShut {NoStop}%
\bibitem [{\citenamefont {Cohen}\ \emph {et~al.}(2006)\citenamefont {Cohen},
  \citenamefont {Davidovitch}, \citenamefont {Schofield}, \citenamefont
  {Brenner},\ and\ \citenamefont {Weitz}}]{ISI:000242219900019}%
  \BibitemOpen
  \bibfield  {author} {\bibinfo {author} {\bibfnamefont {I.}~\bibnamefont
  {Cohen}}, \bibinfo {author} {\bibfnamefont {B.}~\bibnamefont {Davidovitch}},
  \bibinfo {author} {\bibfnamefont {A.~B.}\ \bibnamefont {Schofield}}, \bibinfo
  {author} {\bibfnamefont {M.~P.}\ \bibnamefont {Brenner}}, \ and\ \bibinfo
  {author} {\bibfnamefont {D.~A.}\ \bibnamefont {Weitz}},\ }\href {\doibase
  {10.1103/PhysRevLett.97.215502}} {\bibfield  {journal} {\bibinfo  {journal}
  {{Phys. Rev. Lett.}}\ }\textbf {\bibinfo {volume} {{97}}},\ \bibinfo {pages}
  {215502} (\bibinfo {year} {{2006}})}\BibitemShut {NoStop}%
\bibitem [{\citenamefont {Perge}\ \emph {et~al.}(2014)\citenamefont {Perge},
  \citenamefont {Taberlet}, \citenamefont {Gibaud},\ and\ \citenamefont
  {Manneville}}]{ISI:000342206200016}%
  \BibitemOpen
  \bibfield  {author} {\bibinfo {author} {\bibfnamefont {C.}~\bibnamefont
  {Perge}}, \bibinfo {author} {\bibfnamefont {N.}~\bibnamefont {Taberlet}},
  \bibinfo {author} {\bibfnamefont {T.}~\bibnamefont {Gibaud}}, \ and\ \bibinfo
  {author} {\bibfnamefont {S.}~\bibnamefont {Manneville}},\ }\href {\doibase
  {10.1122/1.4887081}} {\bibfield  {journal} {\bibinfo  {journal} {{J.
  Rheol.}}\ }\textbf {\bibinfo {volume} {{58}}},\ \bibinfo {pages} {1331}
  (\bibinfo {year} {{2014}})}\BibitemShut {NoStop}%
\bibitem [{\citenamefont {Guo}\ \emph {et~al.}(2011)\citenamefont {Guo},
  \citenamefont {Yu}, \citenamefont {Xu},\ and\ \citenamefont
  {Zhou}}]{ISI:000288162500028}%
  \BibitemOpen
  \bibfield  {author} {\bibinfo {author} {\bibfnamefont {Y.}~\bibnamefont
  {Guo}}, \bibinfo {author} {\bibfnamefont {W.}~\bibnamefont {Yu}}, \bibinfo
  {author} {\bibfnamefont {Y.}~\bibnamefont {Xu}}, \ and\ \bibinfo {author}
  {\bibfnamefont {C.}~\bibnamefont {Zhou}},\ }\href {\doibase
  {10.1039/c0sm00970a}} {\bibfield  {journal} {\bibinfo  {journal} {{Soft
  Matter}}\ }\textbf {\bibinfo {volume} {{7}}},\ \bibinfo {pages} {2433}
  (\bibinfo {year} {{2011}})}\BibitemShut {NoStop}%
\bibitem [{\citenamefont {Dimitriou}\ \emph {et~al.}(2013)\citenamefont
  {Dimitriou}, \citenamefont {Ewoldt},\ and\ \citenamefont
  {McKinley}}]{ISI:000312240900002}%
  \BibitemOpen
  \bibfield  {author} {\bibinfo {author} {\bibfnamefont {C.~J.}\ \bibnamefont
  {Dimitriou}}, \bibinfo {author} {\bibfnamefont {R.~H.}\ \bibnamefont
  {Ewoldt}}, \ and\ \bibinfo {author} {\bibfnamefont {G.~H.}\ \bibnamefont
  {McKinley}},\ }\href {\doibase {10.1122/1.4754023}} {\bibfield  {journal}
  {\bibinfo  {journal} {{J. Rheol.}}\ }\textbf {\bibinfo {volume} {{57}}},\
  \bibinfo {pages} {27} (\bibinfo {year} {{2013}})}\BibitemShut {NoStop}%
\bibitem [{\citenamefont {Gibaud}\ \emph {et~al.}(2016)\citenamefont {Gibaud},
  \citenamefont {Perge}, \citenamefont {Lindstrom}, \citenamefont {Taberlet},\
  and\ \citenamefont {Manneville}}]{ISI:000369750400007}%
  \BibitemOpen
  \bibfield  {author} {\bibinfo {author} {\bibfnamefont {T.}~\bibnamefont
  {Gibaud}}, \bibinfo {author} {\bibfnamefont {C.}~\bibnamefont {Perge}},
  \bibinfo {author} {\bibfnamefont {S.~B.}\ \bibnamefont {Lindstrom}}, \bibinfo
  {author} {\bibfnamefont {N.}~\bibnamefont {Taberlet}}, \ and\ \bibinfo
  {author} {\bibfnamefont {S.}~\bibnamefont {Manneville}},\ }\href {\doibase
  {10.1039/c5sm02587g}} {\bibfield  {journal} {\bibinfo  {journal} {{Soft
  Matter}}\ }\textbf {\bibinfo {volume} {{12}}},\ \bibinfo {pages} {1701}
  (\bibinfo {year} {{2016}})}\BibitemShut {NoStop}%
\bibitem [{\citenamefont {Park}\ and\ \citenamefont
  {Rogers}(2018)}]{ISI:000438883200004}%
  \BibitemOpen
  \bibfield  {author} {\bibinfo {author} {\bibfnamefont {J.~D.}\ \bibnamefont
  {Park}}\ and\ \bibinfo {author} {\bibfnamefont {S.~A.}\ \bibnamefont
  {Rogers}},\ }\href {\doibase 10.1122/1.5024701} {\bibfield  {journal}
  {\bibinfo  {journal} {Journal of Rheology}\ }\textbf {\bibinfo {volume}
  {62}},\ \bibinfo {pages} {869} (\bibinfo {year} {2018})}\BibitemShut
  {NoStop}%
\bibitem [{\citenamefont {Radhakrishnan}\ and\ \citenamefont
  {Fielding}(2018)}]{ISI:000427032600013}%
  \BibitemOpen
  \bibfield  {author} {\bibinfo {author} {\bibfnamefont {R.}~\bibnamefont
  {Radhakrishnan}}\ and\ \bibinfo {author} {\bibfnamefont {S.~M.}\ \bibnamefont
  {Fielding}},\ }\href {\doibase {10.1122/1.5023381}} {\bibfield  {journal}
  {\bibinfo  {journal} {{J. Rheol.}}\ }\textbf {\bibinfo {volume} {{62}}},\
  \bibinfo {pages} {559} (\bibinfo {year} {{2018}})}\BibitemShut {NoStop}%
\bibitem [{\citenamefont {Radhakrishnan}\ and\ \citenamefont
  {Fielding}(2016)}]{ISI:000390226100005}%
  \BibitemOpen
  \bibfield  {author} {\bibinfo {author} {\bibfnamefont {R.}~\bibnamefont
  {Radhakrishnan}}\ and\ \bibinfo {author} {\bibfnamefont {S.~M.}\ \bibnamefont
  {Fielding}},\ }\href {\doibase {10.1103/PhysRevLett.117.188001}} {\bibfield
  {journal} {\bibinfo  {journal} {{Phys. Rev. Lett.}}\ }\textbf {\bibinfo
  {volume} {{117}}},\ \bibinfo {pages} {188001} (\bibinfo {year}
  {{2016}})}\BibitemShut {NoStop}%
\bibitem [{\citenamefont {BANNANTINE}\ \emph {et~al.}(1990)\citenamefont
  {BANNANTINE}, \citenamefont {COMER},\ and\ \citenamefont
  {HANDROCK}}]{bannantine1990fundamentals}%
  \BibitemOpen
  \bibfield  {author} {\bibinfo {author} {\bibfnamefont {J.}~\bibnamefont
  {BANNANTINE}}, \bibinfo {author} {\bibfnamefont {J.}~\bibnamefont {COMER}}, \
  and\ \bibinfo {author} {\bibfnamefont {J.}~\bibnamefont {HANDROCK}},\
  }\href@noop {} {\bibfield  {journal} {\bibinfo  {journal} {Research supported
  by the University of Illinois. Englewood Cliffs, NJ, Prentice Hall, 1990,
  286}\ } (\bibinfo {year} {1990})}\BibitemShut {NoStop}%
\bibitem [{\citenamefont {Karmakar}\ \emph {et~al.}(2010)\citenamefont
  {Karmakar}, \citenamefont {Lerner},\ and\ \citenamefont
  {Procaccia}}]{ISI:000280777900001}%
  \BibitemOpen
  \bibfield  {author} {\bibinfo {author} {\bibfnamefont {S.}~\bibnamefont
  {Karmakar}}, \bibinfo {author} {\bibfnamefont {E.}~\bibnamefont {Lerner}}, \
  and\ \bibinfo {author} {\bibfnamefont {I.}~\bibnamefont {Procaccia}},\ }\href
  {\doibase {10.1103/PhysRevE.82.026104}} {\bibfield  {journal} {\bibinfo
  {journal} {{Phys. Rev. E}}\ }\textbf {\bibinfo {volume} {{82}}},\ \bibinfo
  {pages} {026104} (\bibinfo {year} {{2010}})}\BibitemShut {NoStop}%
\bibitem [{\citenamefont {Radhakrishnan}\ \emph {et~al.}(2017)\citenamefont
  {Radhakrishnan}, \citenamefont {Divoux}, \citenamefont {Manneville},\ and\
  \citenamefont {Fielding}}]{ISI:000396030800011}%
  \BibitemOpen
  \bibfield  {author} {\bibinfo {author} {\bibfnamefont {R.}~\bibnamefont
  {Radhakrishnan}}, \bibinfo {author} {\bibfnamefont {T.}~\bibnamefont
  {Divoux}}, \bibinfo {author} {\bibfnamefont {S.}~\bibnamefont {Manneville}},
  \ and\ \bibinfo {author} {\bibfnamefont {S.~M.}\ \bibnamefont {Fielding}},\
  }\href {\doibase {10.1039/c6sm02581a}} {\bibfield  {journal} {\bibinfo
  {journal} {{Soft Matter}}\ }\textbf {\bibinfo {volume} {{13}}},\ \bibinfo
  {pages} {1834} (\bibinfo {year} {{2017}})}\BibitemShut {NoStop}%
\bibitem [{\citenamefont {Divoux}\ \emph {et~al.}(2013)\citenamefont {Divoux},
  \citenamefont {Grenard},\ and\ \citenamefont
  {Manneville}}]{ISI:000313006100057}%
  \BibitemOpen
  \bibfield  {author} {\bibinfo {author} {\bibfnamefont {T.}~\bibnamefont
  {Divoux}}, \bibinfo {author} {\bibfnamefont {V.}~\bibnamefont {Grenard}}, \
  and\ \bibinfo {author} {\bibfnamefont {S.}~\bibnamefont {Manneville}},\
  }\href {\doibase {10.1103/PhysRevLett.110.018304}} {\bibfield  {journal}
  {\bibinfo  {journal} {{Phys. Rev. Lett.}}\ }\textbf {\bibinfo {volume}
  {{110}}},\ \bibinfo {pages} {018304} (\bibinfo {year} {{2013}})}\BibitemShut
  {NoStop}%
\bibitem [{\citenamefont {Colombo}\ and\ \citenamefont
  {Del~Gado}(2014)}]{ISI:000342206200002}%
  \BibitemOpen
  \bibfield  {author} {\bibinfo {author} {\bibfnamefont {J.}~\bibnamefont
  {Colombo}}\ and\ \bibinfo {author} {\bibfnamefont {E.}~\bibnamefont
  {Del~Gado}},\ }\href {\doibase {10.1122/1.4882021}} {\bibfield  {journal}
  {\bibinfo  {journal} {{J. Rheol.}}\ }\textbf {\bibinfo {volume} {{58}}},\
  \bibinfo {pages} {1089} (\bibinfo {year} {{2014}})}\BibitemShut {NoStop}%
\bibitem [{\citenamefont {Shrivastav}\ \emph {et~al.}(2016)\citenamefont
  {Shrivastav}, \citenamefont {Chaudhuri},\ and\ \citenamefont
  {Horbach}}]{ISI:000386386400004}%
  \BibitemOpen
  \bibfield  {author} {\bibinfo {author} {\bibfnamefont {G.~P.}\ \bibnamefont
  {Shrivastav}}, \bibinfo {author} {\bibfnamefont {P.}~\bibnamefont
  {Chaudhuri}}, \ and\ \bibinfo {author} {\bibfnamefont {J.}~\bibnamefont
  {Horbach}},\ }\href {\doibase {10.1103/PhysRevE.94.042605}} {\bibfield
  {journal} {\bibinfo  {journal} {{Phys. Rev. E}}\ }\textbf {\bibinfo {volume}
  {{94}}},\ \bibinfo {pages} {042605} (\bibinfo {year} {{2016}})}\BibitemShut
  {NoStop}%
\bibitem [{\citenamefont {Shi}\ \emph {et~al.}(2007)\citenamefont {Shi},
  \citenamefont {Katz}, \citenamefont {Li},\ and\ \citenamefont
  {Falk}}]{ISI:000246210200042}%
  \BibitemOpen
  \bibfield  {author} {\bibinfo {author} {\bibfnamefont {Y.}~\bibnamefont
  {Shi}}, \bibinfo {author} {\bibfnamefont {M.~B.}\ \bibnamefont {Katz}},
  \bibinfo {author} {\bibfnamefont {H.}~\bibnamefont {Li}}, \ and\ \bibinfo
  {author} {\bibfnamefont {M.~L.}\ \bibnamefont {Falk}},\ }\href {\doibase
  {10.1103/PhysRevLett.98.185505}} {\bibfield  {journal} {\bibinfo  {journal}
  {{Phys. Rev. Lett.}}\ }\textbf {\bibinfo {volume} {{98}}},\ \bibinfo {pages}
  {185505} (\bibinfo {year} {{2007}})}\BibitemShut {NoStop}%
\bibitem [{\citenamefont {Varnik}\ \emph {et~al.}(2004)\citenamefont {Varnik},
  \citenamefont {Bocquet},\ and\ \citenamefont
  {Barrat}}]{varnik-jcp-120-2788-2004}%
  \BibitemOpen
  \bibfield  {author} {\bibinfo {author} {\bibfnamefont {F.}~\bibnamefont
  {Varnik}}, \bibinfo {author} {\bibfnamefont {L.}~\bibnamefont {Bocquet}}, \
  and\ \bibinfo {author} {\bibfnamefont {J.~L.}\ \bibnamefont {Barrat}},\
  }\href@noop {} {\bibfield  {journal} {\bibinfo  {journal} {J. Chem. Phys.}\
  }\textbf {\bibinfo {volume} {120}},\ \bibinfo {pages} {2788} (\bibinfo {year}
  {2004})}\BibitemShut {NoStop}%
\bibitem [{\citenamefont {Kabla}\ \emph {et~al.}(2007)\citenamefont {Kabla},
  \citenamefont {Scheibert},\ and\ \citenamefont
  {Debregeas}}]{ISI:000250675300003}%
  \BibitemOpen
  \bibfield  {author} {\bibinfo {author} {\bibfnamefont {A.}~\bibnamefont
  {Kabla}}, \bibinfo {author} {\bibfnamefont {J.}~\bibnamefont {Scheibert}}, \
  and\ \bibinfo {author} {\bibfnamefont {G.}~\bibnamefont {Debregeas}},\ }\href
  {\doibase {10.1017/S0022112007007276}} {\bibfield  {journal} {\bibinfo
  {journal} {{J. Fluid Mech}}\ }\textbf {\bibinfo {volume} {{587}}},\ \bibinfo
  {pages} {45} (\bibinfo {year} {{2007}})}\BibitemShut {NoStop}%
\bibitem [{\citenamefont {Barry}\ \emph {et~al.}(2010)\citenamefont {Barry},
  \citenamefont {Weaire},\ and\ \citenamefont {Hutzler}}]{ISI:000278158800014}%
  \BibitemOpen
  \bibfield  {author} {\bibinfo {author} {\bibfnamefont {J.~D.}\ \bibnamefont
  {Barry}}, \bibinfo {author} {\bibfnamefont {D.}~\bibnamefont {Weaire}}, \
  and\ \bibinfo {author} {\bibfnamefont {S.}~\bibnamefont {Hutzler}},\ }\href
  {\doibase {10.1007/s00397-010-0449-6}} {\bibfield  {journal} {\bibinfo
  {journal} {{Rheol. Acta}}\ }\textbf {\bibinfo {volume} {{49}}},\ \bibinfo
  {pages} {687} (\bibinfo {year} {{2010}})}\BibitemShut {NoStop}%
\bibitem [{\citenamefont {Moorcroft}\ \emph {et~al.}(2011)\citenamefont
  {Moorcroft}, \citenamefont {Cates},\ and\ \citenamefont
  {Fielding}}]{ISI:000286879900011}%
  \BibitemOpen
  \bibfield  {author} {\bibinfo {author} {\bibfnamefont {R.~L.}\ \bibnamefont
  {Moorcroft}}, \bibinfo {author} {\bibfnamefont {M.~E.}\ \bibnamefont
  {Cates}}, \ and\ \bibinfo {author} {\bibfnamefont {S.~M.}\ \bibnamefont
  {Fielding}},\ }\href {\doibase {10.1103/PhysRevLett.106.055502}} {\bibfield
  {journal} {\bibinfo  {journal} {{Phys. Rev. Lett.}}\ }\textbf {\bibinfo
  {volume} {{106}}},\ \bibinfo {pages} {055502} (\bibinfo {year}
  {{2011}})}\BibitemShut {NoStop}%
\bibitem [{\citenamefont {Fielding}(2014)}]{ISI:000344142000001}%
  \BibitemOpen
  \bibfield  {author} {\bibinfo {author} {\bibfnamefont {S.~M.}\ \bibnamefont
  {Fielding}},\ }\href {\doibase {10.1088/0034-4885/77/10/102601}} {\bibfield
  {journal} {\bibinfo  {journal} {{Rep. Prog. Phys.}}\ }\textbf {\bibinfo
  {volume} {{77}}},\ \bibinfo {pages} {102601} (\bibinfo {year}
  {{2014}})}\BibitemShut {NoStop}%
\bibitem [{\citenamefont {Lehtinen}\ \emph {et~al.}(2013)\citenamefont
  {Lehtinen}, \citenamefont {Puisto}, \citenamefont {Illa}, \citenamefont
  {Mohtaschemi},\ and\ \citenamefont {Alava}}]{ISI:000322544200022}%
  \BibitemOpen
  \bibfield  {author} {\bibinfo {author} {\bibfnamefont {A.}~\bibnamefont
  {Lehtinen}}, \bibinfo {author} {\bibfnamefont {A.}~\bibnamefont {Puisto}},
  \bibinfo {author} {\bibfnamefont {X.}~\bibnamefont {Illa}}, \bibinfo {author}
  {\bibfnamefont {M.}~\bibnamefont {Mohtaschemi}}, \ and\ \bibinfo {author}
  {\bibfnamefont {M.~J.}\ \bibnamefont {Alava}},\ }\href {\doibase
  {10.1039/c3sm50988e}} {\bibfield  {journal} {\bibinfo  {journal} {{Soft
  Matter}}\ }\textbf {\bibinfo {volume} {{9}}},\ \bibinfo {pages} {8041}
  (\bibinfo {year} {{2013}})}\BibitemShut {NoStop}%
\bibitem [{\citenamefont {Manning}\ \emph {et~al.}(2007)\citenamefont
  {Manning}, \citenamefont {Langer},\ and\ \citenamefont
  {Carlson}}]{PhysRevE.76.056106}%
  \BibitemOpen
  \bibfield  {author} {\bibinfo {author} {\bibfnamefont {M.~L.}\ \bibnamefont
  {Manning}}, \bibinfo {author} {\bibfnamefont {J.~S.}\ \bibnamefont {Langer}},
  \ and\ \bibinfo {author} {\bibfnamefont {J.~M.}\ \bibnamefont {Carlson}},\
  }\href {\doibase 10.1103/PhysRevE.76.056106} {\bibfield  {journal} {\bibinfo
  {journal} {Phys. Rev. E}\ }\textbf {\bibinfo {volume} {76}},\ \bibinfo
  {pages} {056106} (\bibinfo {year} {2007})}\BibitemShut {NoStop}%
\bibitem [{\citenamefont {Manning}\ \emph {et~al.}(2009)\citenamefont
  {Manning}, \citenamefont {Daub}, \citenamefont {Langer},\ and\ \citenamefont
  {Carlson}}]{Manningetal2009a}%
  \BibitemOpen
  \bibfield  {author} {\bibinfo {author} {\bibfnamefont {M.~L.}\ \bibnamefont
  {Manning}}, \bibinfo {author} {\bibfnamefont {E.~G.}\ \bibnamefont {Daub}},
  \bibinfo {author} {\bibfnamefont {J.~S.}\ \bibnamefont {Langer}}, \ and\
  \bibinfo {author} {\bibfnamefont {J.~M.}\ \bibnamefont {Carlson}},\ }\href
  {\doibase 10.1103/PhysRevE.79.016110} {\bibfield  {journal} {\bibinfo
  {journal} {Phys. Rev. E}\ }\textbf {\bibinfo {volume} {79}},\ \bibinfo
  {pages} {016110} (\bibinfo {year} {2009})}\BibitemShut {NoStop}%
\bibitem [{\citenamefont {Jagla}(2010)}]{ISI:000285583500029}%
  \BibitemOpen
  \bibfield  {author} {\bibinfo {author} {\bibfnamefont {E.~A.}\ \bibnamefont
  {Jagla}},\ }\href {\doibase {10.1088/1742-5468/2010/12/P12025}} {\bibfield
  {journal} {\bibinfo  {journal} {{J. Stat. Mech. Theory E}}\ }\textbf
  {\bibinfo {volume} {{2010}}},\ \bibinfo {pages} {P12025} (\bibinfo {year}
  {{2010}})}\BibitemShut {NoStop}%
\bibitem [{\citenamefont {Chaudhuri}\ and\ \citenamefont
  {Horbach}(2013)}]{ISI:000325376600001}%
  \BibitemOpen
  \bibfield  {author} {\bibinfo {author} {\bibfnamefont {P.}~\bibnamefont
  {Chaudhuri}}\ and\ \bibinfo {author} {\bibfnamefont {J.}~\bibnamefont
  {Horbach}},\ }\href {\doibase {10.1103/PhysRevE.88.040301}} {\bibfield
  {journal} {\bibinfo  {journal} {{Phys. Rev. E}}\ }\textbf {\bibinfo {volume}
  {{88}}},\ \bibinfo {pages} {040301} (\bibinfo {year} {{2013}})}\BibitemShut
  {NoStop}%
\bibitem [{\citenamefont {Moorcroft}\ and\ \citenamefont
  {Fielding}(2013)}]{ISI:000315141600016}%
  \BibitemOpen
  \bibfield  {author} {\bibinfo {author} {\bibfnamefont {R.~L.}\ \bibnamefont
  {Moorcroft}}\ and\ \bibinfo {author} {\bibfnamefont {S.~M.}\ \bibnamefont
  {Fielding}},\ }\href {\doibase {10.1103/PhysRevLett.110.086001}} {\bibfield
  {journal} {\bibinfo  {journal} {{Phys. Rev. Lett.}}\ }\textbf {\bibinfo
  {volume} {{110}}},\ \bibinfo {pages} {086001} (\bibinfo {year}
  {{2013}})}\BibitemShut {NoStop}%
\bibitem [{\citenamefont {Coussot}\ \emph {et~al.}(2002)\citenamefont
  {Coussot}, \citenamefont {Raynaud}, \citenamefont {Bertrand}, \citenamefont
  {Moucheront}, \citenamefont {Guilbaud}, \citenamefont {Huynh}, \citenamefont
  {Jarny},\ and\ \citenamefont {Lesueur}}]{Coussotetal2002c}%
  \BibitemOpen
  \bibfield  {author} {\bibinfo {author} {\bibfnamefont {P.}~\bibnamefont
  {Coussot}}, \bibinfo {author} {\bibfnamefont {J.~S.}\ \bibnamefont
  {Raynaud}}, \bibinfo {author} {\bibfnamefont {F.}~\bibnamefont {Bertrand}},
  \bibinfo {author} {\bibfnamefont {P.}~\bibnamefont {Moucheront}}, \bibinfo
  {author} {\bibfnamefont {J.~P.}\ \bibnamefont {Guilbaud}}, \bibinfo {author}
  {\bibfnamefont {H.~T.}\ \bibnamefont {Huynh}}, \bibinfo {author}
  {\bibfnamefont {S.}~\bibnamefont {Jarny}}, \ and\ \bibinfo {author}
  {\bibfnamefont {D.}~\bibnamefont {Lesueur}},\ }\href {\doibase
  10.1103/PhysRevLett.88.218301} {\bibfield  {journal} {\bibinfo  {journal}
  {Phys. Rev. Lett.}\ }\textbf {\bibinfo {volume} {88}},\ \bibinfo {pages}
  {218301} (\bibinfo {year} {2002})}\BibitemShut {NoStop}%
\bibitem [{\citenamefont {Ragouilliaux}\ \emph {et~al.}(2006)\citenamefont
  {Ragouilliaux}, \citenamefont {Herzhaft}, \citenamefont {Bertrand},\ and\
  \citenamefont {Coussot}}]{ISI:000242151800010}%
  \BibitemOpen
  \bibfield  {author} {\bibinfo {author} {\bibfnamefont {A.}~\bibnamefont
  {Ragouilliaux}}, \bibinfo {author} {\bibfnamefont {B.}~\bibnamefont
  {Herzhaft}}, \bibinfo {author} {\bibfnamefont {F.}~\bibnamefont {Bertrand}},
  \ and\ \bibinfo {author} {\bibfnamefont {P.}~\bibnamefont {Coussot}},\ }\href
  {\doibase {10.1007/s00397-006-0114-2}} {\bibfield  {journal} {\bibinfo
  {journal} {{Rheol. Acta}}\ }\textbf {\bibinfo {volume} {{46}}},\ \bibinfo
  {pages} {261} (\bibinfo {year} {{2006}})}\BibitemShut {NoStop}%
\bibitem [{\citenamefont {Brader}\ \emph {et~al.}(2008)\citenamefont {Brader},
  \citenamefont {Cates},\ and\ \citenamefont {Fuchs}}]{ISI:000259680600077}%
  \BibitemOpen
  \bibfield  {author} {\bibinfo {author} {\bibfnamefont {J.~M.}\ \bibnamefont
  {Brader}}, \bibinfo {author} {\bibfnamefont {M.~E.}\ \bibnamefont {Cates}}, \
  and\ \bibinfo {author} {\bibfnamefont {M.}~\bibnamefont {Fuchs}},\ }\href
  {\doibase {10.1103/PhysRevLett.101.138301}} {\bibfield  {journal} {\bibinfo
  {journal} {{Phys. Rev. Lett.}}\ }\textbf {\bibinfo {volume} {{101}}},\
  \bibinfo {pages} {138301} (\bibinfo {year} {{2008}})}\BibitemShut {NoStop}%
\bibitem [{\citenamefont {Amann}\ \emph {et~al.}(2013)\citenamefont {Amann},
  \citenamefont {Siebenbuerger}, \citenamefont {Krueger}, \citenamefont
  {Weysser}, \citenamefont {Ballauff},\ and\ \citenamefont
  {Fuchs}}]{ISI:000312240900007}%
  \BibitemOpen
  \bibfield  {author} {\bibinfo {author} {\bibfnamefont {C.~P.}\ \bibnamefont
  {Amann}}, \bibinfo {author} {\bibfnamefont {M.}~\bibnamefont
  {Siebenbuerger}}, \bibinfo {author} {\bibfnamefont {M.}~\bibnamefont
  {Krueger}}, \bibinfo {author} {\bibfnamefont {F.}~\bibnamefont {Weysser}},
  \bibinfo {author} {\bibfnamefont {M.}~\bibnamefont {Ballauff}}, \ and\
  \bibinfo {author} {\bibfnamefont {M.}~\bibnamefont {Fuchs}},\ }\href
  {\doibase {10.1122/1.4764000}} {\bibfield  {journal} {\bibinfo  {journal}
  {{J. Rheol.}}\ }\textbf {\bibinfo {volume} {{57}}},\ \bibinfo {pages} {149}
  (\bibinfo {year} {{2013}})}\BibitemShut {NoStop}%
\bibitem [{\citenamefont {Papenkort}\ and\ \citenamefont
  {Voigtmann}(2015)}]{ISI:000366319700022}%
  \BibitemOpen
  \bibfield  {author} {\bibinfo {author} {\bibfnamefont {S.}~\bibnamefont
  {Papenkort}}\ and\ \bibinfo {author} {\bibfnamefont {T.}~\bibnamefont
  {Voigtmann}},\ }\href {\doibase {10.1063/1.4936358}} {\bibfield  {journal}
  {\bibinfo  {journal} {{J. Chem. Phys.}}\ }\textbf {\bibinfo {volume}
  {{143}}},\ \bibinfo {pages} {204502} (\bibinfo {year} {{2015}})}\BibitemShut
  {NoStop}%
\bibitem [{\citenamefont {Nicolas}\ \emph {et~al.}(2018)\citenamefont
  {Nicolas}, \citenamefont {Ferrero}, \citenamefont {Martens},\ and\
  \citenamefont {Barrat}}]{nicolas2018deformation}%
  \BibitemOpen
  \bibfield  {author} {\bibinfo {author} {\bibfnamefont {A.}~\bibnamefont
  {Nicolas}}, \bibinfo {author} {\bibfnamefont {E.~E.}\ \bibnamefont
  {Ferrero}}, \bibinfo {author} {\bibfnamefont {K.}~\bibnamefont {Martens}}, \
  and\ \bibinfo {author} {\bibfnamefont {J.-L.}\ \bibnamefont {Barrat}},\
  }\href@noop {} {\bibfield  {journal} {\bibinfo  {journal} {Rev. Mod. Phys.}\
  }\textbf {\bibinfo {volume} {90}},\ \bibinfo {pages} {045006} (\bibinfo
  {year} {2018})}\BibitemShut {NoStop}%
\bibitem [{\citenamefont {Picard}\ \emph {et~al.}(2004)\citenamefont {Picard},
  \citenamefont {Ajdari}, \citenamefont {Lequeux},\ and\ \citenamefont
  {Bocquet}}]{ISI:000226675600003}%
  \BibitemOpen
  \bibfield  {author} {\bibinfo {author} {\bibfnamefont {G.}~\bibnamefont
  {Picard}}, \bibinfo {author} {\bibfnamefont {A.}~\bibnamefont {Ajdari}},
  \bibinfo {author} {\bibfnamefont {F.}~\bibnamefont {Lequeux}}, \ and\
  \bibinfo {author} {\bibfnamefont {L.}~\bibnamefont {Bocquet}},\ }\href
  {\doibase {10.1140/epje/i2004-10054-8}} {\bibfield  {journal} {\bibinfo
  {journal} {{Eur. Phys. J. E}}\ }\textbf {\bibinfo {volume} {{15}}},\ \bibinfo
  {pages} {371} (\bibinfo {year} {{2004}})}\BibitemShut {NoStop}%
\bibitem [{\citenamefont {H{\'e}braud}\ and\ \citenamefont
  {Lequeux}(1998)}]{hebraud1998mode}%
  \BibitemOpen
  \bibfield  {author} {\bibinfo {author} {\bibfnamefont {P.}~\bibnamefont
  {H{\'e}braud}}\ and\ \bibinfo {author} {\bibfnamefont {F.}~\bibnamefont
  {Lequeux}},\ }\href@noop {} {\bibfield  {journal} {\bibinfo  {journal}
  {Physical review letters}\ }\textbf {\bibinfo {volume} {81}},\ \bibinfo
  {pages} {2934} (\bibinfo {year} {1998})}\BibitemShut {NoStop}%
\bibitem [{\citenamefont {Lin}\ and\ \citenamefont
  {Wyart}(2016)}]{ISI:000368519600002}%
  \BibitemOpen
  \bibfield  {author} {\bibinfo {author} {\bibfnamefont {J.}~\bibnamefont
  {Lin}}\ and\ \bibinfo {author} {\bibfnamefont {M.}~\bibnamefont {Wyart}},\
  }\href {\doibase {10.1103/PhysRevX.6.011005}} {\bibfield  {journal} {\bibinfo
   {journal} {{Phys. Rev. X}}\ }\textbf {\bibinfo {volume} {{6}}},\ \bibinfo
  {pages} {011005} (\bibinfo {year} {{2016}})}\BibitemShut {NoStop}%
\bibitem [{\citenamefont {Fielding}\ \emph {et~al.}(2009)\citenamefont
  {Fielding}, \citenamefont {Cates},\ and\ \citenamefont
  {Sollich}}]{ISI:000266798200007}%
  \BibitemOpen
  \bibfield  {author} {\bibinfo {author} {\bibfnamefont {S.~M.}\ \bibnamefont
  {Fielding}}, \bibinfo {author} {\bibfnamefont {M.~E.}\ \bibnamefont {Cates}},
  \ and\ \bibinfo {author} {\bibfnamefont {P.}~\bibnamefont {Sollich}},\ }\href
  {\doibase {10.1039/b812394m}} {\bibfield  {journal} {\bibinfo  {journal}
  {{Soft Matter}}\ }\textbf {\bibinfo {volume} {{5}}},\ \bibinfo {pages} {2378}
  (\bibinfo {year} {{2009}})}\BibitemShut {NoStop}%
\bibitem [{\citenamefont {Hoyle}\ and\ \citenamefont
  {Fielding}(2015)}]{ISI:000352990500012}%
  \BibitemOpen
  \bibfield  {author} {\bibinfo {author} {\bibfnamefont {D.~M.}\ \bibnamefont
  {Hoyle}}\ and\ \bibinfo {author} {\bibfnamefont {S.~M.}\ \bibnamefont
  {Fielding}},\ }\href {\doibase {10.1103/PhysRevLett.114.158301}} {\bibfield
  {journal} {\bibinfo  {journal} {{Phys. Rev. Lett.}}\ }\textbf {\bibinfo
  {volume} {{114}}},\ \bibinfo {pages} {158301} (\bibinfo {year}
  {{2015}})}\BibitemShut {NoStop}%
\bibitem [{\citenamefont {Papanastasiou}(1987)}]{ISI:A1987H891000002}%
  \BibitemOpen
  \bibfield  {author} {\bibinfo {author} {\bibfnamefont {T.~C.}\ \bibnamefont
  {Papanastasiou}},\ }\href {\doibase {10.1122/1.549926}} {\bibfield  {journal}
  {\bibinfo  {journal} {{J. Rheol.}}\ }\textbf {\bibinfo {volume} {{31}}},\
  \bibinfo {pages} {385} (\bibinfo {year} {{1987}})}\BibitemShut {NoStop}%
\bibitem [{\citenamefont {Syrakos}\ \emph {et~al.}(2014)\citenamefont
  {Syrakos}, \citenamefont {Georgiou},\ and\ \citenamefont
  {Alexandrou}}]{ISI:000337874400008}%
  \BibitemOpen
  \bibfield  {author} {\bibinfo {author} {\bibfnamefont {A.}~\bibnamefont
  {Syrakos}}, \bibinfo {author} {\bibfnamefont {G.~C.}\ \bibnamefont
  {Georgiou}}, \ and\ \bibinfo {author} {\bibfnamefont {A.~N.}\ \bibnamefont
  {Alexandrou}},\ }\href {\doibase {10.1016/j.jnnfm.2014.03.004}} {\bibfield
  {journal} {\bibinfo  {journal} {{J. Non-Newton. Fluid.}}\ }\textbf {\bibinfo
  {volume} {{208}}},\ \bibinfo {pages} {88} (\bibinfo {year}
  {{2014}})}\BibitemShut {NoStop}%
\bibitem [{\citenamefont {Liu}\ \emph {et~al.}(2002)\citenamefont {Liu},
  \citenamefont {Muller},\ and\ \citenamefont {Denn}}]{ISI:000173617600006}%
  \BibitemOpen
  \bibfield  {author} {\bibinfo {author} {\bibfnamefont {B.}~\bibnamefont
  {Liu}}, \bibinfo {author} {\bibfnamefont {S.}~\bibnamefont {Muller}}, \ and\
  \bibinfo {author} {\bibfnamefont {M.}~\bibnamefont {Denn}},\ }\href {\doibase
  {10.1016/S0377-0257(01)00177-X}} {\bibfield  {journal} {\bibinfo  {journal}
  {{J. Non-Newton. Fluid.}}\ }\textbf {\bibinfo {volume} {{102}}},\ \bibinfo
  {pages} {179} (\bibinfo {year} {{2002}})}\BibitemShut {NoStop}%
\bibitem [{\citenamefont {Beris}\ \emph {et~al.}(1985)\citenamefont {Beris},
  \citenamefont {Tsamopoulos}, \citenamefont {Armstrong},\ and\ \citenamefont
  {Brown}}]{ISI:A1985ATS6200011}%
  \BibitemOpen
  \bibfield  {author} {\bibinfo {author} {\bibfnamefont {A.~N.}\ \bibnamefont
  {Beris}}, \bibinfo {author} {\bibfnamefont {J.~A.}\ \bibnamefont
  {Tsamopoulos}}, \bibinfo {author} {\bibfnamefont {R.~C.}\ \bibnamefont
  {Armstrong}}, \ and\ \bibinfo {author} {\bibfnamefont {R.~A.}\ \bibnamefont
  {Brown}},\ }\href {\doibase {10.1017/S0022112085002622}} {\bibfield
  {journal} {\bibinfo  {journal} {{J. Fluid Mech}}\ }\textbf {\bibinfo {volume}
  {{158}}},\ \bibinfo {pages} {219} (\bibinfo {year} {{1985}})}\BibitemShut
  {NoStop}%
\bibitem [{\citenamefont {Smyrnaios}\ and\ \citenamefont
  {Tsamopoulos}(2001)}]{ISI:000171099500010}%
  \BibitemOpen
  \bibfield  {author} {\bibinfo {author} {\bibfnamefont {D.}~\bibnamefont
  {Smyrnaios}}\ and\ \bibinfo {author} {\bibfnamefont {J.}~\bibnamefont
  {Tsamopoulos}},\ }\href {\doibase {10.1016/S0377-0257(01)00141-0}} {\bibfield
   {journal} {\bibinfo  {journal} {{J. Non-Newton. Fluid.}}\ }\textbf {\bibinfo
  {volume} {{100}}},\ \bibinfo {pages} {165} (\bibinfo {year}
  {{2001}})}\BibitemShut {NoStop}%
\bibitem [{\citenamefont {Blackery}\ and\ \citenamefont
  {Mitsoulis}(1997)}]{ISI:A1997XB14900003}%
  \BibitemOpen
  \bibfield  {author} {\bibinfo {author} {\bibfnamefont {J.}~\bibnamefont
  {Blackery}}\ and\ \bibinfo {author} {\bibfnamefont {E.}~\bibnamefont
  {Mitsoulis}},\ }\href {\doibase {10.1016/S0377-0257(96)01536-4}} {\bibfield
  {journal} {\bibinfo  {journal} {{J. Non-Newton. Fluid.}}\ }\textbf {\bibinfo
  {volume} {{70}}},\ \bibinfo {pages} {59} (\bibinfo {year}
  {{1997}})}\BibitemShut {NoStop}%
\bibitem [{\citenamefont {Mitsoulis}\ and\ \citenamefont
  {Tsamopoulos}(2017)}]{ISI:000398072300007}%
  \BibitemOpen
  \bibfield  {author} {\bibinfo {author} {\bibfnamefont {E.}~\bibnamefont
  {Mitsoulis}}\ and\ \bibinfo {author} {\bibfnamefont {J.}~\bibnamefont
  {Tsamopoulos}},\ }\href {\doibase {10.1007/s00397-016-0981-0}} {\bibfield
  {journal} {\bibinfo  {journal} {{Rheol. Acta}}\ }\textbf {\bibinfo {volume}
  {{56}}},\ \bibinfo {pages} {231} (\bibinfo {year} {{2017}})}\BibitemShut
  {NoStop}%
\bibitem [{\citenamefont {Keshavarz}\ \emph {et~al.}(2017)\citenamefont
  {Keshavarz}, \citenamefont {Divoux}, \citenamefont {Manneville},\ and\
  \citenamefont {McKinley}}]{ISI:000406087600005}%
  \BibitemOpen
  \bibfield  {author} {\bibinfo {author} {\bibfnamefont {B.}~\bibnamefont
  {Keshavarz}}, \bibinfo {author} {\bibfnamefont {T.}~\bibnamefont {Divoux}},
  \bibinfo {author} {\bibfnamefont {S.}~\bibnamefont {Manneville}}, \ and\
  \bibinfo {author} {\bibfnamefont {G.~H.}\ \bibnamefont {McKinley}},\ }\href
  {\doibase {10.1021/acsmacrolett.7b00213}} {\bibfield  {journal} {\bibinfo
  {journal} {{ACS Macro Lett.}}\ }\textbf {\bibinfo {volume} {{6}}},\ \bibinfo
  {pages} {663} (\bibinfo {year} {{2017}})}\BibitemShut {NoStop}%
\bibitem [{\citenamefont {Jaishankar}\ and\ \citenamefont
  {McKinley}(2014)}]{ISI:000345194700006}%
  \BibitemOpen
  \bibfield  {author} {\bibinfo {author} {\bibfnamefont {A.}~\bibnamefont
  {Jaishankar}}\ and\ \bibinfo {author} {\bibfnamefont {G.~H.}\ \bibnamefont
  {McKinley}},\ }\href {\doibase {10.1122/1.4892114}} {\bibfield  {journal}
  {\bibinfo  {journal} {{J. Rheol.}}\ }\textbf {\bibinfo {volume} {{58}}},\
  \bibinfo {pages} {1751} (\bibinfo {year} {{2014}})}\BibitemShut {NoStop}%
\bibitem [{\citenamefont {Wei}\ \emph {et~al.}(2018)\citenamefont {Wei},
  \citenamefont {Solomon},\ and\ \citenamefont {Larson}}]{ISI:000419395900024}%
  \BibitemOpen
  \bibfield  {author} {\bibinfo {author} {\bibfnamefont {Y.}~\bibnamefont
  {Wei}}, \bibinfo {author} {\bibfnamefont {M.~J.}\ \bibnamefont {Solomon}}, \
  and\ \bibinfo {author} {\bibfnamefont {R.~G.}\ \bibnamefont {Larson}},\
  }\href {\doibase {10.1122/1.4996752}} {\bibfield  {journal} {\bibinfo
  {journal} {{J. Rheol.}}\ }\textbf {\bibinfo {volume} {{62}}},\ \bibinfo
  {pages} {321} (\bibinfo {year} {{2018}})}\BibitemShut {NoStop}%
\bibitem [{\citenamefont {Wei}\ \emph {et~al.}(2016)\citenamefont {Wei},
  \citenamefont {Solomon},\ and\ \citenamefont {Larson}}]{ISI:000388430800020}%
  \BibitemOpen
  \bibfield  {author} {\bibinfo {author} {\bibfnamefont {Y.}~\bibnamefont
  {Wei}}, \bibinfo {author} {\bibfnamefont {M.~J.}\ \bibnamefont {Solomon}}, \
  and\ \bibinfo {author} {\bibfnamefont {R.~G.}\ \bibnamefont {Larson}},\
  }\href {\doibase {10.1122/1.4965228}} {\bibfield  {journal} {\bibinfo
  {journal} {{J. Rheol.}}\ }\textbf {\bibinfo {volume} {{60}}},\ \bibinfo
  {pages} {1301} (\bibinfo {year} {{2016}})}\BibitemShut {NoStop}%
\bibitem [{\citenamefont {Armstrong}\ \emph {et~al.}(2017)\citenamefont
  {Armstrong}, \citenamefont {Beris}, \citenamefont {Rogers},\ and\
  \citenamefont {Wagner}}]{ISI:000410747200004}%
  \BibitemOpen
  \bibfield  {author} {\bibinfo {author} {\bibfnamefont {M.~J.}\ \bibnamefont
  {Armstrong}}, \bibinfo {author} {\bibfnamefont {A.~N.}\ \bibnamefont
  {Beris}}, \bibinfo {author} {\bibfnamefont {S.~A.}\ \bibnamefont {Rogers}}, \
  and\ \bibinfo {author} {\bibfnamefont {N.~J.}\ \bibnamefont {Wagner}},\
  }\href {\doibase {10.1007/s00397-017-1038-8}} {\bibfield  {journal} {\bibinfo
   {journal} {{Rheol. Acta}}\ }\textbf {\bibinfo {volume} {{56}}},\ \bibinfo
  {pages} {811} (\bibinfo {year} {{2017}})}\BibitemShut {NoStop}%
\bibitem [{\citenamefont {Saramito}(2007)}]{ISI:000249049200001}%
  \BibitemOpen
  \bibfield  {author} {\bibinfo {author} {\bibfnamefont {P.}~\bibnamefont
  {Saramito}},\ }\href {\doibase {10.1016/j.jnnfm.2007.04.004}} {\bibfield
  {journal} {\bibinfo  {journal} {{J. Non-Newton. Fluid.}}\ }\textbf {\bibinfo
  {volume} {{145}}},\ \bibinfo {pages} {1} (\bibinfo {year}
  {{2007}})}\BibitemShut {NoStop}%
\bibitem [{\citenamefont {de~Souza~Mendes}(2011)}]{ISI:000288162500032}%
  \BibitemOpen
  \bibfield  {author} {\bibinfo {author} {\bibfnamefont {P.~R.}\ \bibnamefont
  {de~Souza~Mendes}},\ }\href {\doibase {10.1039/c0sm01021a}} {\bibfield
  {journal} {\bibinfo  {journal} {{Soft Matter}}\ }\textbf {\bibinfo {volume}
  {{7}}},\ \bibinfo {pages} {2471} (\bibinfo {year} {{2011}})}\BibitemShut
  {NoStop}%
\bibitem [{\citenamefont {Saramito}(2009)}]{ISI:000265317500017}%
  \BibitemOpen
  \bibfield  {author} {\bibinfo {author} {\bibfnamefont {P.}~\bibnamefont
  {Saramito}},\ }\href {\doibase {10.1016/j.jnnfm.2008.12.001}} {\bibfield
  {journal} {\bibinfo  {journal} {{J. Nonnewton. Fluid Mech.}}\ }\textbf
  {\bibinfo {volume} {{158}}},\ \bibinfo {pages} {154} (\bibinfo {year}
  {{2009}})}\BibitemShut {NoStop}%
\bibitem [{\citenamefont {Patinet}\ \emph {et~al.}(2016)\citenamefont
  {Patinet}, \citenamefont {Vandembroucq},\ and\ \citenamefont
  {Falk}}]{ISI:000380122800005}%
  \BibitemOpen
  \bibfield  {author} {\bibinfo {author} {\bibfnamefont {S.}~\bibnamefont
  {Patinet}}, \bibinfo {author} {\bibfnamefont {D.}~\bibnamefont
  {Vandembroucq}}, \ and\ \bibinfo {author} {\bibfnamefont {M.~L.}\
  \bibnamefont {Falk}},\ }\href {\doibase {10.1103/PhysRevLett.117.045501}}
  {\bibfield  {journal} {\bibinfo  {journal} {{Phys. Rev. Lett.}}\ }\textbf
  {\bibinfo {volume} {{117}}},\ \bibinfo {pages} {045501} (\bibinfo {year}
  {{2016}})}\BibitemShut {NoStop}%
\bibitem [{\citenamefont {Barbot}\ \emph {et~al.}(2018)\citenamefont {Barbot},
  \citenamefont {Lerbinger}, \citenamefont {Hernandez-Garcia}, \citenamefont
  {Garcia-Garcia}, \citenamefont {Falk}, \citenamefont {Vandembroucq},\ and\
  \citenamefont {Patinet}}]{ISI:000426631500004}%
  \BibitemOpen
  \bibfield  {author} {\bibinfo {author} {\bibfnamefont {A.}~\bibnamefont
  {Barbot}}, \bibinfo {author} {\bibfnamefont {M.}~\bibnamefont {Lerbinger}},
  \bibinfo {author} {\bibfnamefont {A.}~\bibnamefont {Hernandez-Garcia}},
  \bibinfo {author} {\bibfnamefont {R.}~\bibnamefont {Garcia-Garcia}}, \bibinfo
  {author} {\bibfnamefont {M.~L.}\ \bibnamefont {Falk}}, \bibinfo {author}
  {\bibfnamefont {D.}~\bibnamefont {Vandembroucq}}, \ and\ \bibinfo {author}
  {\bibfnamefont {S.}~\bibnamefont {Patinet}},\ }\href {\doibase
  {10.1103/PhysRevE.97.033001}} {\bibfield  {journal} {\bibinfo  {journal}
  {{Phys. Rev. E}}\ }\textbf {\bibinfo {volume} {{97}}},\ \bibinfo {pages}
  {033001} (\bibinfo {year} {{2018}})}\BibitemShut {NoStop}%
\bibitem [{Rot()}]{RottlerPreprint}%
  \BibitemOpen
  \href@noop {} {}\bibinfo {note} {C. Ruscher and J. Rottler,
  arxiv.org/pdf/1908.01081.pdf}\BibitemShut {NoStop}%
\bibitem [{Sol()}]{SollichCecam}%
  \BibitemOpen
  \href@noop {} {}\bibinfo {note} {P. Sollich, in CECAM Workshop (ACAM, Dublin,
  Ireland, 2011).}\BibitemShut {Stop}%
\bibitem [{\citenamefont {Bouchaud}\ and\ \citenamefont
  {Dean}(1995)}]{bouchaud1995aging}%
  \BibitemOpen
  \bibfield  {author} {\bibinfo {author} {\bibfnamefont {J.-P.}\ \bibnamefont
  {Bouchaud}}\ and\ \bibinfo {author} {\bibfnamefont {D.~S.}\ \bibnamefont
  {Dean}},\ }\href@noop {} {\bibfield  {journal} {\bibinfo  {journal} {Journal
  de Physique I}\ }\textbf {\bibinfo {volume} {5}},\ \bibinfo {pages} {265}
  (\bibinfo {year} {1995})}\BibitemShut {NoStop}%
\bibitem [{\citenamefont {Cates}\ and\ \citenamefont
  {Sollich}(2004)}]{ISI:000187952400012}%
  \BibitemOpen
  \bibfield  {author} {\bibinfo {author} {\bibfnamefont {M.}~\bibnamefont
  {Cates}}\ and\ \bibinfo {author} {\bibfnamefont {P.}~\bibnamefont
  {Sollich}},\ }\href {\doibase {10.1122/1.1634985}} {\bibfield  {journal}
  {\bibinfo  {journal} {{J. Rheol.}}\ }\textbf {\bibinfo {volume} {{48}}},\
  \bibinfo {pages} {193} (\bibinfo {year} {{2004}})}\BibitemShut {NoStop}%
\bibitem [{\citenamefont {Purnomo}\ \emph {et~al.}(2008)\citenamefont
  {Purnomo}, \citenamefont {van~den Ende}, \citenamefont {Vanapalli},\ and\
  \citenamefont {Mugele}}]{purnomo2008glass}%
  \BibitemOpen
  \bibfield  {author} {\bibinfo {author} {\bibfnamefont {E.~H.}\ \bibnamefont
  {Purnomo}}, \bibinfo {author} {\bibfnamefont {D.}~\bibnamefont {van~den
  Ende}}, \bibinfo {author} {\bibfnamefont {S.~A.}\ \bibnamefont {Vanapalli}},
  \ and\ \bibinfo {author} {\bibfnamefont {F.}~\bibnamefont {Mugele}},\
  }\href@noop {} {\bibfield  {journal} {\bibinfo  {journal} {Physical review
  letters}\ }\textbf {\bibinfo {volume} {101}},\ \bibinfo {pages} {238301}
  (\bibinfo {year} {2008})}\BibitemShut {NoStop}%
\bibitem [{\citenamefont {Rogers}\ \emph {et~al.}(2011)\citenamefont {Rogers},
  \citenamefont {Erwin}, \citenamefont {Vlassopoulos},\ and\ \citenamefont
  {Cloitre}}]{ISI:000287095800012}%
  \BibitemOpen
  \bibfield  {author} {\bibinfo {author} {\bibfnamefont {S.~A.}\ \bibnamefont
  {Rogers}}, \bibinfo {author} {\bibfnamefont {B.~M.}\ \bibnamefont {Erwin}},
  \bibinfo {author} {\bibfnamefont {D.}~\bibnamefont {Vlassopoulos}}, \ and\
  \bibinfo {author} {\bibfnamefont {M.}~\bibnamefont {Cloitre}},\ }\href
  {\doibase {10.1122/1.3544591}} {\bibfield  {journal} {\bibinfo  {journal}
  {{J. Rheol.}}\ }\textbf {\bibinfo {volume} {{55}}},\ \bibinfo {pages} {435}
  (\bibinfo {year} {{2011}})}\BibitemShut {NoStop}%
\bibitem [{\citenamefont {Bauschinger}(1886)}]{bauschinger1886veranderung}%
  \BibitemOpen
  \bibfield  {author} {\bibinfo {author} {\bibfnamefont {J.}~\bibnamefont
  {Bauschinger}},\ }\href@noop {} {\bibfield  {journal} {\bibinfo  {journal}
  {Mitteilungen des mechanisch-technischen Laboratoriums der K{\"o}niglich
  Technischen Hochschule M{\"u}nchen}\ }\textbf {\bibinfo {volume} {13}}
  (\bibinfo {year} {1886})}\BibitemShut {NoStop}%
\bibitem [{\citenamefont {Liu}\ \emph {et~al.}(2018)\citenamefont {Liu},
  \citenamefont {Martens},\ and\ \citenamefont {Barrat}}]{ISI:000423131200025}%
  \BibitemOpen
  \bibfield  {author} {\bibinfo {author} {\bibfnamefont {C.}~\bibnamefont
  {Liu}}, \bibinfo {author} {\bibfnamefont {K.}~\bibnamefont {Martens}}, \ and\
  \bibinfo {author} {\bibfnamefont {J.-L.}\ \bibnamefont {Barrat}},\ }\href
  {\doibase {10.1103/PhysRevLett.120.028004}} {\bibfield  {journal} {\bibinfo
  {journal} {{Phys. Rev. Lett.}}\ }\textbf {\bibinfo {volume} {{120}}},\
  \bibinfo {pages} {028004} (\bibinfo {year} {{2018}})}\BibitemShut {NoStop}%
\end{thebibliography}

%

\end{document}